\documentclass[aps,prd,preprint,eqsecnum,tightenlines,nofootinbib,showpacs]{revtex4}
\usepackage[dvips]{graphicx}
\usepackage{epsf,subfigure}
\usepackage{amsfonts}
\newif\ifdraft
\drafttrue
\newif\ifpreprint
\preprinttrue

\def\fig#1{Fig.~{\ref{#1}}}

\def\eqn#1{eq.~({\ref{#1}})}
\def\eqns#1#2{eqs.~({\ref{#1}}) and~({\ref{#2}})} 
\def\sect#1{Section~{\ref{#1}}}

\def\tab#1{Table~{\ref{#1}}}

\def\App#1{Appendix~{\ref{#1}}}
\def\MHVbar{{$\overline{\hbox{MHV}}$}}
\def\nn{\nonumber}
\def\NeqFour{\mathcal{N}=4}

\def\NeqEight{\mathcal{N}=8}

\def\I{{\cal I}}
\def\bowtie{{\rm bow\mbox{-}tie}}
\def\P{{\rm P}}
\def\NP{{\rm NP}}

\def\tree{{\rm tree}}
\def\oneloop{{1 \mbox{-} \rm loop}}

\def\mud{\lambda}
\def\n{\tilde n}
\def\f{\tilde f}
\def\pol{\varepsilon}

\def\eps{\epsilon}

\def\tree{{\rm tree}}

\def\Tr{{\rm Tr}}

\def\spa#1.#2{\left\langle#1\,#2\right\rangle}
\def\spb#1.#2{\left[#1\,#2\right]}

\newbox\charbox
\newbox\slabox
\def\s#1{{      
        \setbox\charbox=\hbox{$#1$}
        \setbox\slabox=\hbox{$/$}
        \dimen\charbox=\ht\slabox
        \advance\dimen\charbox by -\dp\slabox
        \advance\dimen\charbox by -\ht\charbox
        \advance\dimen\charbox by \dp\charbox
        \divide\dimen\charbox by 2
        \raise-\dimen\charbox\hbox to \wd\charbox{\hss/\hss}
        \llap{$#1$}
}}

\newskip\humongous \humongous=0pt plus 1000pt minus 100pt

\newif\ifdtup

\newcounter{eqnumber}[section]

\begin{document}

\title{
\ifpreprint
 \hbox{\normalsize \rm MCTP-12-21 \hskip 10 cm   UCLA/12/TEP/106} 
\hbox{$\null$}
\fi
\Large Ultraviolet Cancellations in Half-Maximal 
Supergravity as a Consequence 
of the\\ Double-Copy Structure
}
 
\author{Zvi~Bern$^a$, Scott~Davies$^a$, Tristan~Dennen$^{b}$
 and Yu-tin~Huang$^{c}$}

\affiliation{
$a$ Department of Physics and Astronomy, University of California 
at Los Angeles\\ 
 Los Angeles, CA 90095-1547, USA \\ 
$\null$ \\
$b$ Niels Bohr International Academy and Discovery Center\\
The Niels Bohr Institute\\
Blegdamsvej 17, DK-2100 Copenhagen, Denmark\\
$\null$ \\
$c$ Michigan Center for Theoretical Physics \\
Randall Laboratory of Physics\\
University of Michigan, Ann Arbor, MI 48109, USA
$\null$ \\
}

\vskip .5 cm
\begin{abstract}
We show that the double-copy structure of gravity forbids divergences
in pure half-maximal (16 supercharge) supergravity at four and five
points at one loop in $D<8$ and at two loops in $D<6$.  We link the
cancellations that render these supergravity amplitudes finite to
corresponding ones that eliminate forbidden color factors from the
divergences of pure nonsupersymmetric Yang-Mills theory.  The
vanishing of the two-loop four-point divergence in $D=5$ half-maximal
supergravity is an example where a valid counterterm satisfying the
known symmetries exists, yet is not present.  We also give explicit
forms of divergences in half-maximal supergravity at one loop in $D=8$
and at two loops in $D=6$.
\end{abstract}

\pacs{04.65.+e, 11.15.Bt, 11.30.Pb, 11.55.Bq \hspace{1cm}}

\maketitle

\section{Introduction}

Recent years have made it clear that even at loop level perturbative
scattering amplitudes in gravity theories are closely related to
corresponding ones in gauge theories.  In particular, a recent
conjecture holds that whenever a duality between color and kinematics
is made manifest, the integrands of (super)gravity loop amplitudes can
be obtained immediately from corresponding gauge-theory
ones~\cite{BCJ, BCJLoop}. It has been clear since the original
loop-level double-copy construction that it would have important
implications for resolving long-standing questions on the ultraviolet
properties of gravity theories.  An obvious question is whether it can
be used to show that $\NeqEight$~\cite{N8Sugra} and other supergravity
theories have a tamer than expected ultraviolet behavior.  If each
order of the perturbative expansion were finite, it would imply a
deep new structure of the theory.

The double-copy structure has been used to simplify new nontrivial
calculations of the ultraviolet properties of supergravity amplitudes,
demonstrating behavior remarkably similar to corresponding
gauge-theory amplitudes. In explicit calculations of amplitudes one
can, of course, directly confirm that the conjectured duality and
double-copy properties hold.  For example, through at least four
loops, the ultraviolet divergences of $\NeqEight$ supergravity in the
critical dimension where they first occur are proportional to
divergences appearing in the subleading-color terms of corresponding
$\NeqFour$ super-Yang-Mills
amplitudes~\cite{CompactThree,BCJ5Point,ck4l}.  The double-copy
construction also played a key role in a recent computation showing
that all three-loop four-point amplitudes in $\NeqFour$ supergravity
in $D=4$~\cite{N4Sugra} are ultraviolet finite~\cite{N4grav}, contrary
to expectations based on the availability of an apparently
supersymmetric and duality invariant $R^4$
counterterm~\cite{VanishingVolume}.

In this paper, we use the double-copy construction to explain
ultraviolet finiteness in a simpler example: the four- and five-point
two-loop potential divergences in $D=5$ half-maximal
supergravity~\cite{GrisaruSiegel}.  We directly link the finiteness of
the half-maximal supergravity four- and five-point amplitudes at one
loop in $D<8$ and at two loops in $D<6$ to ultraviolet cancellations
of forbidden color factors in gauge-theory amplitudes.  This can be
understood in terms of generalized gauge invariance~\cite{BCJ,
  BCJLoop, BCJSquare, Tye}, which links the symmetries and
cancellations of gauge theory to those of gravity.  We note that the
absence of the potential $D=5$ two-loop four-point divergence has been
seen from string-theory calculations as well, so it offers a good way 
to expose cancellations in the theory~\cite{VanhoveN4}.

At present there does not appear to be an argument restricting
counterterms using the conventional symmetries of the theory to rule
out this divergence.  Indeed as shown in
ref.~\cite{BossardHoweStelle5D} the counterterm appears to be
expressible as a duality invariant full superspace integral of a
density (which itself is not duality invariant). It would be very
important to fully understand the extent to which duality symmetry and
supersymmetry by themselves can shed light on counterterm restrictions in
half-maximal supergravity at two loops in $D=5$.

Some cases we study here are especially simple to analyze
because the $\NeqFour$ super-Yang-Mills amplitudes used on one side of
the double-copy construction have diagrammatic numerators that are
independent of loop momenta.  Because of this property, even after
performing the loop integration, the corresponding amplitudes in pure
supergravity theories with sixteen or more supercharges are simple
linear combinations of corresponding gauge-theory
amplitudes~\cite{OneLoopN4,TwoLoopN4}.  Indeed, using the double-copy
construction, in ref.~\cite{Schnitzer} one-loop four- and five-point
and two-loop four-point gravity amplitudes were expressed directly in
terms of certain subleading-color amplitudes of corresponding gauge
theories. There the authors found cancellations leading to the relatively mild
infrared singularities in gravity, similar to the way we find tamer
ultraviolet behavior in gravity than in the gauge-theory amplitudes
from which they are built.

Besides the double-copy relation between gravity and gauge theory, there are
other reasons to believe that the ultraviolet behavior of gravity
might be better than expected from applying standard symmetry
arguments.  Even pure Einstein gravity at one loop exhibits remarkable
cancellations as the number of external legs increases, essentially
scaling with the number of external legs in the same way as gauge
theory~\cite{UnexpectedOneLoop,DunbarEttleI}.  Through unitarity, such
cancellations feed into nontrivial ultraviolet cancellations at all
loop orders~\cite{Finite}.  Very recently, resummations of ${\cal N}
\ge 4$ supergravity amplitudes were shown to have surprisingly good
behavior in the high-energy Regge limit~\cite{LipatovGravity},
suggestive of a connection to the surprisingly good ultraviolet
behavior of loop amplitudes in these theories.

Whether the observed cancellations are sufficient to render the theory
ultraviolet finite remains an open question.  (For a recent optimistic
opinion in favor of ultraviolet finiteness of $\NeqEight$ supergravity
see ref.~\cite{Kallosh}. For a recent pessimistic opinion see
ref.~\cite{Banks}.)  In $\NeqEight$ supergravity in $D=4$, in
particular, no divergence can occur before seven loops, but a
consensus holds that a valid $D^8R^4$ counterterm exists at seven
loops~\cite{SevenLoopGravity}.  This may seem to suggest that in $D=4$
the theory diverges at seven loops~\cite{SevenLoopGravity}.
Interestingly, the candidate full-superspace integral for the
counterterm turns out to vanish~\cite{VanishingVolume}, leaving only a
BPS candidate counterterm represented by an integral over $7/8$ of the
superspace. The potential three-loop counterterm of $\NeqFour$
supergravity in $D=4$~\cite{N4Sugra} is analogous in this regard, as
it too is BPS.  In a previous paper~\cite{N4grav}, we proved by direct
computation that the coefficient of the expected three-loop
counterterm in $\NeqFour$ supergravity vanishes.  (See
ref.~\cite{VanhoveN4} for a string-theoretic argument of this
vanishing and ref.~\cite{RenataSergioN4} for a conjecture linking it
to a hidden superconformal invariance.)  While no nonrenormalization
theorems are known for these cases, an important open question remains
whether the BPS nature of the counterterm plays a role in explaining
the finiteness.  In any case, based on the vanishing of divergences in
explicit calculations presented here and in ref.~\cite{N4grav}, we see
that arguments based on applying the known symmetries of supergravity
theories can be misleading.  It is therefore important to carry out
explicit computations to guide future studies.  In particular, the
arguments suggesting a seven-loop divergence in $D=4$ also suggest
that in higher dimensions, $\NeqEight$ supergravity will be worse
behaved than $\NeqFour$ super-Yang-Mills theory starting at five loops
due to the availability of a $D^8 R^4$ counterterm.  It should be
possible to test this by direct computation~\cite{Neq54np}.

Besides explaining the nontrivial cancellation of two-loop four-point
divergences in half-maximal supergravity in $D=5$, we also present the
explicit forms of one-loop four- and five-point divergences in $D=8$
and two-loop four-point divergences in $D=6$. We obtain these using
the same double-copy construction as used to demonstrate the vanishing
of all three-loop divergences of $\NeqFour$ supergravity in
$D=4$~\cite{N4grav}.  In this construction, one copy is a maximally
supersymmetric Yang-Mills amplitude in a form in which the duality
between color and kinematics holds manifestly~\cite{BCJLoop}, while
the second copy uses ordinary Feynman rules in Feynman gauge.  The
diagrams are then expanded for large loop momenta (or equivalently
small external momenta) and integrated to extract the ultraviolet
divergences~\cite{MarcusSagnotti}.  The explicit expressions for
divergences presented here should be useful in future studies of the
symmetries and structure of half-maximal supergravity.

This paper is organized as follows.  In \sect{ReviewSection}, we
briefly review some basic features of the duality between color and
kinematics and the double-copy construction of gravity.  In
\sect{OneloopUVSection}, we show that at four and five points the
potential one-loop divergences in half-maximal supergravity cancel in $D<8$ by
linking them to forbidden divergences in corresponding gauge-theory
amplitudes.  We also present the explicit form of one-loop
divergences in $D=8$. Then in \sect{TwoloopUVSection} we show that the
two-loop four- and five-point amplitudes of half-maximal supergravity do not
have divergences in $D<6$. In addition, this section contains an explicit
expression for four-point $D=6$ divergences. We give our conclusions and
outlook in \sect{ConclusionSection}.  An appendix computing the
two-loop four-point divergence of pure Yang-Mills theory in $D=5$ is
also included.  These results are used in \sect{TwoloopUVSection} to
explicitly demonstrate ultraviolet cancellations in the corresponding
half-maximal supergravity amplitude.

\section{Review of BCJ Duality}
\label{ReviewSection}

In this section we review the duality between color and kinematics
conjectured by Carrasco, Johansson and one of the authors (BCJ) and the
related double-copy construction of gravity loop
amplitudes~\cite{BCJ,BCJLoop}.  These properties underlie our ability
to analyze the divergence structure of half-maximal supergravity
amplitudes.  Recent applications to the half-maximal theory of
$\NeqFour$ supergravity in $D=4$ can be found in
refs.~\cite{OneLoopN4,TwoLoopN4,N4grav,Schnitzer}.

\subsection{Duality between color and kinematics}

We can write any $m$-point $L$-loop gauge-theory amplitude with
all particles in the adjoint representation as
\begin{equation}
 {\cal A}^{L-\rm loop}_m =  {i^L} {g^{m-2 +2L }}
\sum_{{\cal S}_m} \sum_{j}{\int \prod_{l=1}^L \frac{d^{D} p_l}{ (2 \pi)^{D}}
  \frac{1}{S_j}  \frac {n_j c_j}{\prod_{\alpha_j}{p^2_{\alpha_j}}}}\,.
\label{LoopGauge} 
\end{equation}
The sum labeled by $j$ runs over the set of distinct non-isomorphic
$m$-point $L$-loop graphs with only cubic (i.e. trivalent)
vertices. $S_j$ is the symmetry factor of graph $j$, removing
overcounts from the sum over $m!$ permutations of external legs
indicated by ${\cal S}_m$ and from internal automorphism symmetry.
The product in the denominator runs over all Feynman propagators of
graph $j$.  The integrals are over $L$ independent $D$-dimensional
loop momenta.  The $c_j$ are the color factors obtained by dressing
every three-vertex with a group-theory structure constant,
\begin{equation}
\f^{abc} = i \sqrt{2} f^{abc}=\Tr([T^{a},T^{b}]T^{c}) \,,
\label{fabcdef}
\end{equation}
and $n_j$ are kinematic numerators of graph $j$ depending on momenta,
polarizations and spinors.  For supersymmetric amplitudes expressed in
superspace, there will also be Grassmann parameters in the numerators.
Contact terms in the amplitude are expressed in this form by
multiplying and dividing by appropriate propagators.  We note that there is enormous freedom in the
choice of numerators, due to generalized gauge
invariance~\cite{BCJ, BCJLoop, BCJSquare, Tye}.

The  conjectured duality of refs.~\cite{BCJ,BCJLoop} states that to all
loop orders there should exist a form of the amplitude where kinematic
numerators satisfy the same algebraic relations as color factors.
For Yang-Mills theory this amounts to imposing the same 
Jacobi identities on the kinematic numerators as satisfied
by the color factors,
\begin{equation}
c_i = c_j - c_k \;  \Rightarrow \;  n_i = n_j - n_k \,,
\label{BCJDuality}
\end{equation}
where the indices $i,j,k$ denote the diagram to which the color
factors and numerators belong.  Moreover, the numerator factors are
required to have the same antisymmetry property as color factors under
interchange of two legs attaching to a cubic vertex,
\begin{equation}
c_i \rightarrow - c_i \;  \Rightarrow \;  n_i \rightarrow - n_i \,.
\label{BCJAntiSymmetry}
\end{equation}
As explained in some detail in refs.~\cite{HenrikJJReview,
  ck4l,BCJVirtuousTrees}, the numerator relations are functional
equations.  For four-point tree amplitudes such relations were noticed
long ago~\cite{Halzen}. Beyond the four-point tree level, the
relations are rather nontrivial and work only after appropriate
rearrangements of the amplitudes.

At tree level, explicit forms of amplitudes satisfying the duality
have been found for an arbitrary number of external
legs~\cite{TreeAllN}.  An interesting consequence of the duality is
that color-ordered partial tree amplitudes satisfy nontrivial
relations~\cite{BCJ}. These have been proven both in gauge theory and
in string theory~\cite{BCJProofs}.  The duality is natural to
understand using the heterotic string because of the parallel
treatment of color and kinematics~\cite{Tye}.  Although we do not yet
have a satisfactory Lagrangian understanding, some progress in this
direction can be found in refs.~\cite{BCJSquare,OConnell1}.  The
duality (\ref{BCJDuality}) has also been expressed in terms of an
alternative trace-based representation~\cite{Trace}, emphasizing the
underlying group-theoretic structure of the duality.  Indeed, progress
has been made in understanding the underlying infinite-dimensional Lie
algebra~\cite{OConnell1,OConnell2}.  Interestingly, the duality
between color and kinematics also appears to hold in more exotic
three-dimensional theories~\cite{BLGTheory}, as well as in certain
cases with higher-dimension operators~\cite{BroedelDixon}.  Relations
similar to tree-level ones have also been shown to hold for the
identical helicity one-loop amplitudes of pure Yang-Mills
theory~\cite{OneloopBCJLike}.

At loop level, although the duality remains a conjecture, a number of
nontrivial checks have been carried out.  The duality has been
confirmed to hold up to four loops for the four-point amplitudes of
$\NeqFour$ super-Yang-Mills theory~\cite{BCJLoop,ck4l}, and for the
five-point one- and two-loop amplitudes of this
theory~\cite{BCJ5Point}.  It is also known to hold for the
identical-helicity one- and two-loop four-point amplitudes of pure
Yang-Mills theory~\cite{BCJLoop}.

\subsection{Gravity as a double copy of gauge theory}

Once the gauge-theory amplitudes have been arranged into the
form (\ref{LoopGauge}) where the numerators
satisfy the duality (\ref{BCJDuality}), the corresponding gravity
loop integrands become remarkably simple to obtain~\cite{BCJ,BCJLoop}
via the replacement, 
\begin{equation}
c_i \rightarrow \n_i\,.
\label{ColorSubstitutionRule}
\end{equation}
The $\n_i$ are diagram numerators from a second gauge theory. 
Making the
substitution (\ref{ColorSubstitutionRule}) in \eqn{LoopGauge} gives us
the double-copy form of gravity amplitudes~\cite{BCJ,BCJLoop},
\begin{equation}
{\cal M}^{L-\rm loop}_m =  {i^{L+1}} {\biggl(\frac{\kappa}{2}\biggr)^{m-2+2L}} \,
\sum_{{\cal S}_m} \sum_{j} {\int \prod_{l=1}^L \frac{d^{D} p_l}{(2 \pi)^{D}}
 \frac{1}{S_j} \frac{n_j \n_j}{\prod_{\alpha_j}{p^2_{\alpha_j}}}} \, , 
\hskip .7 cm 
\label{DoubleCopy}
\end{equation}
where ${\cal M}^{L-\rm loop}_m$ are $m$-point $L$-loop gravity
amplitudes.  In the double-copy formula (\ref{DoubleCopy}), only one of
the two sets of numerators $n_j$ or $\n_j$ needs to satisfy the duality
relation (\ref{BCJDuality}).

Here we are interested in half-maximal supergravity in $D>4$
dimensions.  This theory is obtained via the double-copy formula by
taking the direct product of pure nonsupersymmetric Yang-Mills theory
with maximally supersymmetric Yang-Mills theory.  This construction is
the same one used to construct one- and two-loop amplitudes in
$\NeqFour$ supergravity in
$D=4$~\cite{OneLoopN4,TwoLoopN4,N4grav,Schnitzer}.  While the
maximally supersymmetric Yang-Mills theory has exactly the same number
of states as $\NeqFour$ super-Yang-Mills theory does in four
dimensions, the pure nonsupersymmetric Yang-Mills theory used in this
construction has additional states compared to the $D=4$ case.

At tree level, \eqn{DoubleCopy} encodes the
Kawai-Lewellen-Tye~\cite{KLT} relations between gravity and gauge
theory~\cite{BCJ}.  The double-copy formula has been proven at tree
level when the duality~(\ref{BCJDuality}) holds in the corresponding
gauge theories~\cite{BCJSquare}.  It has also been studied in some
detail in a number of cases through four loops in $\NeqEight$
supergravity~\cite{BCJLoop,BCJ5Point,ck4l}, and through three loops in
$\NeqFour$ supergravity~\cite{OneLoopN4,TwoLoopN4,N4grav}.

\subsection{Supergravity with $Q+16$ supercharges at one and two loops}

%
\begin{figure}
\begin{center}
\includegraphics[scale=0.4]{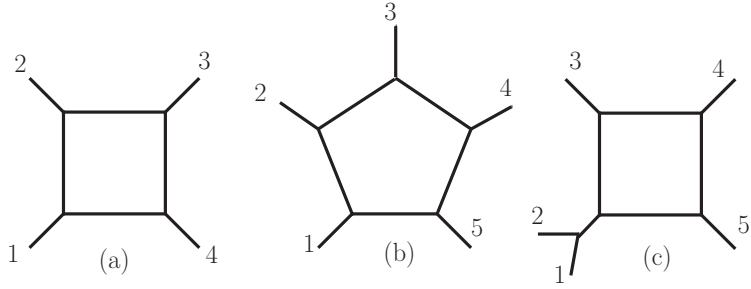}
\caption{Diagram (a) specifies the four-point color factor $c^{(1)}_{1234}$
used in \eqn{OneLoopFourPtGauge}, and 
diagram (b) specifies the color factor $c^{(1)}_{12345}$ in \eqn{OneLoopFivePtGauge}. Diagram (a) and its permutations appear in the four-point amplitude
of maximal super-Yang-Mills theory.  At five points both (b) and (c) and 
their permutations appear.}
\label{OneLoopDiagramsFigure}
\end{center}
\end{figure}

A color-dressed four-point one-loop (super) Yang-Mills amplitude 
can be expressed as~\cite{DixonMaltoniColor}
\begin{eqnarray}
\mathcal{A}^{(1)}_{Q}(1,2,3,4)=g^4\bigg[
  c^{(1)}_{1234}A^{(1)}_{Q}(1,2,3,4)
+ c^{(1)}_{1342}A^{(1)}_{Q}(1,3,4,2)
+ c^{(1)}_{1423}A^{(1)}_{Q}(1,4,2,3)\bigg]\,.
\label{OneLoopFourPtGauge}
\end{eqnarray} 
The $c^{(1)}_{1234}$ are the color factors of a box diagram with
consecutive external legs $(1,2,3,4)$, illustrated in
\fig{OneLoopDiagramsFigure}(a), and dressed with structure constants
$\tilde{f}^{abc}$. Here, $A^{(1)}_{Q}$ are one-loop color-ordered
amplitudes~\cite{ColorLoop}.  The label $Q$ specifies the number of
supercharges. 
For maximally supersymmetric Yang-Mills ($Q=16$), the amplitude is
given by the one-loop scalar box integral, with the corresponding
diagram numerators given by~\cite{GSB}
\begin{equation}
n_{1234} = n_{1342} = n_{1423}= s t A_{Q=16}^\tree (1,2,3,4)\,,
\label{FourPointNumerators}
\end{equation}
where $A_{Q=16}^\tree(1,2,3,4)$ is the color-ordered tree amplitude of
maximal super-Yang-Mills theory in any dimension and for any states of
the theory.  The Mandelstam invariants are defined as $s = (k_1 +
k_2)^2$, $t = (k_2+k_3)^2$ and $u = (k_1 + k_3)^2$.  It is
straightforward to check that this form satisfies the duality between
color and kinematics.

To obtain pure supergravity amplitudes with $Q+16$ supercharges, we
simply replace the color factors with the corresponding numerators
(\ref{FourPointNumerators}), yielding a rather simple formula,
\begin{equation}
\nonumber\mathcal{M}^{(1)}_{Q+16}=
i\bigg(\frac{\kappa}{2}\bigg)^4 s t A^{\rm tree}_{Q=16}(1,2,3,4)
\bigg[A^{(1)}_{Q}(1,2,3,4) + A^{(1)}_{Q}(1,3,4,2)
 + A^{(1)}_{Q}(1,4,2,3) \bigg]\,.
\label{FourPointSupergravity}
\end{equation} 

In four dimensions, the prefactor in \eqn{FourPointSupergravity} can be
written in a supersymmetric form~\cite{Nair},
\begin{equation}
s t  A^{\mathrm{tree}}_{Q=16}(1,2,3,4)=
-i \delta^{(8)}(\mathcal{Q})\, 
\frac{[1\,2][3\,4]}{\langle 1\,2\rangle\langle 3\,4\rangle}\,,
\end{equation}
which makes half the supersymmetries manifest. Here
$\langle 1\,2\rangle$ and $[1\,2]$ are the usual four-dimensional
spinor-inner products for Weyl spinors (see e.g. ref.~\cite{TreeReview}).
In this form all states of the $\NeqFour$ super-Yang-Mills multiplet are
encoded in the Grassmann-valued delta function of the
supercharges $\mathcal{Q}$.  Simple superspace expressions also have
been constructed in six dimensions~\cite{DHS}.  Here we do not use
any superspace properties, other than the fact that all
states are encoded in one simple prefactor.

%
\begin{figure}
\begin{center}
\includegraphics[scale=0.5]{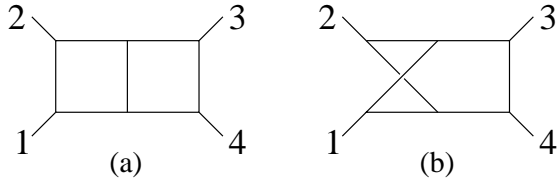}
\caption{The two-loop planar and nonplanar double-box diagrams. }
\label{DoubleBoxFigure}
\end{center}
\end{figure}

The two-loop four-point case is also relatively simple.
The color-dressed two-loop four-point (super) Yang-Mills amplitude can be
conveniently written as~\cite{OneLoopN4,TwoLoopN4}
\begin{eqnarray}
\mathcal{A}^{(2)}_{Q}(1,2,3,4) &=&
g^6\bigg[ c^{\rm P}_{1234} A^{\rm P}_{Q}(1,2,3,4)
+c^{\rm P}_{3421} A^{\rm P}_{Q} (3,4,2,1) \label{YMTwoloop} \\
 && \hskip .5cm   \null 
+ c^{\rm NP}_{1234} A^{\rm NP}_{ Q}(1,2,3,4)
+c^{\rm NP}_{3421} A^{\rm NP}_{Q}(3,4,2,1) + {\rm cyclic}(2,3,4)\bigg]\,, \nn
\end{eqnarray} 
where `cyclic$(2,3,4)$' indicates a sum over the remaining two cyclic
permutations of legs 2, 3 and 4.  Here $c^{\rm P}_{1234}$ and $c^{\rm
  NP}_{1234}$ are the color factors obtained by dressing the planar
and nonplanar double-box diagrams in \fig{DoubleBoxFigure} with
structure constants $\f^{abc}$. The $A^{\rm P}_{Q}$ and 
$A^{\rm NP}_{Q}$ are the integrated planar
and nonplanar kinematic parts of the amplitudes.  The form
(\ref{YMTwoloop}) matches the one used in $\NeqFour$ super-Yang-Mills
theory~\cite{BDDPR}.  This form is valid for any theory with 
only adjoint representation particles, as can be shown using
color Jacobi-identity rearrangements~\cite{DixonMaltoniColor}.

For maximal ($Q=16$) super-Yang-Mills theory in any dimension, the
standard loop-integral representation of the two-loop four-point
amplitude~\cite{BRY,BCJLoop} satisfies the duality between color and
kinematics (\ref{BCJDuality}). An important simplifying feature is
that the numerator factors do not have loop-momentum dependence,
and are
\begin{equation}
n^{\rm P}_{1234} = s^2 t A_{Q=16}^\tree(1,2,3,4)\,, \hskip 2 cm 
n^{\rm NP}_{1234} = s^2 t A_{Q=16}^\tree(1,2,3,4)\,,
\end{equation}
corresponding to the two partial amplitudes $A_{Q=16}^{\rm
  P}(1,2,3,4)$ and $A_{Q=16}^{\rm NP}(1,2,3,4)$ in \eqn{YMTwoloop}.
When constructing gravity amplitudes via the replacement
(\ref{ColorSubstitutionRule}), the numerator of the $\mathcal{N}=4$
super-Yang-Mills copy comes outside the integral, and thus one can
express the integrated supergravity amplitude as a linear combination
of integrated (super) Yang-Mills amplitudes~\cite{OneLoopN4}. Using
this, the integrated four-point two-loop supergravity amplitude
is~\cite{TwoLoopN4}
\begin{eqnarray}
\mathcal{M}^{(2)}_{Q+16}(1,2,3,4) &=&
i\bigg(\frac{\kappa}{2}\bigg)^6stA^{\rm tree}_{Q=16}(1,2,3,4)
\bigg[s \Bigl( A^{\rm P}_{Q}(1,2,3,4) + A^{\rm NP}_{Q}(1,2,3,4) \nn \\
&& \hskip 2 cm \null 
 +  A^{\rm P}_{Q}(3,4,2,1) + A^{\rm NP}_{Q}(3,4,2,1) \Bigr) 
 + {\rm cyclic}(2,3,4)\bigg]\,, \hskip 1 cm 
\label{TwoLoopSugraBCJ}
\end{eqnarray}
which holds in any dimension $D \le 10$ and for pure supergravity theories
with $Q+16$ supercharges.

Now consider the five-point case.  At one loop, a
five-point gauge-theory amplitude with only adjoint representation
particles can be written in the form~\cite{DixonMaltoniColor},
\begin{equation}
{\cal A}_{Q}^{(1)}(1,2,3,4,5) = 
g^5 \sum_{{\cal S}_5/(\mathbb{Z}_5\times \mathbb{Z}_2)} c^{(1)}_{12345} \, 
A_{{Q}}^{(1)}(1,2, 3,4,5)\,,
\label{OneLoopFivePtGauge} 
\end{equation}
where the color factor $c^{(1)}_{12345}$ is that of the pentagon
diagram, displayed in \fig{OneLoopDiagramsFigure}(b). The
sum runs over all permutations with the five cyclic ones and reflections
removed, signified by ${\cal S}_5/(\mathbb{Z}_5\times \mathbb{Z}_2)$.
For the
maximally supersymmetric ($Q=16$) case, only pentagon and box
integrals contribute in the BCJ form~\cite{BCJ5Point}, illustrated
in \fig{OneLoopDiagramsFigure}(b) and (c).

Using the substitution rule (\ref{ColorSubstitutionRule}), together
with the observation that at four and five points, the
duality-satisfying maximal super-Yang-Mills numerators with states
restricted to a four-dimensional subspace are independent of loop
momenta, we immediately obtain the simple expression~\cite{OneLoopN4},
\begin{equation}
{\cal M}_{Q+16}^{(1)}(1,2, 3,4,5) = 
i\Bigl( \frac{\kappa}{2} \Bigr)^5
\sum_{{\cal S}_5/(\mathbb{Z}_5\times \mathbb{Z}_2)}\, \tilde{n}_{12345} \, 
A_{{Q}}^{(1)}(1,2,3,4,5)\,.
\label{SimpleOneloopGravitySusy} 
\end{equation}
For $Q=0$ the obtained amplitudes are those of half-maximal pure
supergravity theory.  

In a four-dimensional external subspace, the 
maximal super-Yang-Mills kinematic numerators appearing in
\eqn{SimpleOneloopGravitySusy} for external gluons in an MHV configuration
are given by~\cite{BCJ5Point}
\begin{equation}
\tilde{n}_{123 45}  = \beta_{12345}  \equiv  \spa{i}.{j}^4
\frac{[1\,2][2\,3][3\,4][4\,5][5\,1]}{4\varepsilon(1,2,3,4)},
\label{OneLoopPentagonNumerators}
\end{equation}
where $i$ and $j$ label the two negative-helicity legs and
$\varepsilon(1,2,3,4)\equiv\varepsilon_{\mu\nu\rho\sigma}k_1^{\mu}k_2^{\nu}k_3^{\rho}k_4^{\sigma}=\mathrm{Det}(k_i^{\mu})$.
The anti-MHV case is given by the parity conjugate.  Five-point
amplitudes with other states beside gluons have also been discussed in
ref.~\cite{BCJ5Point}, but we will not use them here.  A conjectured
$D$-dimensional generalization of these numerator functions may be
found in ref.~\cite{BCJVirtuousTrees}.

While only numerators corresponding to the pentagon diagram in
\fig{OneLoopDiagramsFigure}(b) are required for
\eqn{SimpleOneloopGravitySusy}, in \sect{FivePointOneloopSubsection}
we will use the expressions for the numerators of the box diagrams,
illustrated in \fig{OneLoopDiagramsFigure}(c), as well.  Since the
maximal super-Yang-Mills numerators satisfy the duality
(\ref{BCJDuality}), the box numerators can be written in terms of the pentagon numerators
by the kinematic Jacobi relations: $\tilde{n}_{[12]345} =
\tilde{n}_{12345}-\tilde{n}_{21345}$, where $\tilde{n}_{[12]345}$ is
the numerator for the box diagram in
\fig{OneLoopDiagramsFigure}(c).  This gives
\begin{equation}
\tilde{n}_{[12]345} = \gamma_{12}\equiv\gamma_{12345}\equiv
 \spa{i}.{j}^4
\frac{[1\,2]^2[3\,4][4\,5][3\,5]}{4\varepsilon(1,2,3,4)} \,.
\label{gammas}
\end{equation}
The $\gamma$'s are symmetric in their last three
indices, so they can be specified by the first two indices only.  They
also satisfy the relations,
\begin{equation}
\sum_{i=1}^5\gamma_{ij}=0,\hspace{1cm}\gamma_{ij}=-\gamma_{ji} \,,
\label{gammaBas}
\end{equation}
from which we see that there are six linearly independent $\gamma$'s. 
They are completely interchangeable with the $\beta$'s because, 
\begin{eqnarray}
\gamma_{12}&=&\beta_{12345}-\beta_{21345}, \nonumber \\
\beta_{12345}&=&
\frac{1}{2}\left(\gamma_{12}+\gamma_{13}+\gamma_{14}+\gamma_{23}
   +\gamma_{24}+\gamma_{34}\right) \,,
\end{eqnarray}
so there are also six linearly independent $\beta$'s.

\section{Ultraviolet structure of half-maximal supergravity at one loop}
\label{OneloopUVSection}

In this section, we illustrate how the double copy links cancellations
of supergravity divergences to those of forbidden color factors in
gauge-theory divergences using simple one-loop examples.  In
particular, we discuss the divergence properties of the four- and
five-point amplitudes in higher dimensions from this vantage point.
The one-loop four- and five-point double-copy formulas
(\ref{FourPointSupergravity}) and (\ref{SimpleOneloopGravitySusy})
give integrated supergravity amplitudes with 16 or more supercharges
directly in terms of corresponding integrated (super) Yang-Mills amplitudes.
This allows us to obtain the divergences of these supergravity
amplitudes simply by plugging in known Yang-Mills counterterm
amplitudes.

We note that in $D=4$ the one-loop amplitudes of $\mathcal{N}<8$
supergravity theories have been extensively studied recently in
refs.~\cite{DunbarEttleI,DunbarEttleII}.  For the cases of four and
five points, a double-copy construction has been given in
ref.~\cite{OneLoopN4}.  Very recently the one-loop four-graviton
amplitude for $\mathcal{N}=4$ supergravity coupled to $\mathcal{N}=4$
vector multiplets has also been obtained by taking the field-theory
limit of string-theory results~\cite{OneloopN4Vanhove}.  Here we are
mainly interested in higher dimensions.

\subsection{Four-point divergences at one loop}
\label{FourPointOneLoopSubsection}

We now demonstrate that pure half-maximal supergravity does not have
four-point divergences at one loop for $D < 8$.
In dimensional regularization at one loop, there
can be no divergences in any dimension other than even integer
dimensions.  We will start with a warm up in $D=4$ before turning to
the more interesting cases of $D=6$ and $D=8$.

\subsubsection{$D=4$ warm up}

We start by reproducing the well-known result that the four-point amplitude
of pure $\NeqFour$ supergravity has no divergence at one
loop~\cite{PeterVan}.  The renormalizability
of Yang-Mills theory in $D=4$ implies that the full one-loop
divergence must be proportional to the color-dressed tree amplitude:
\begin{equation}
\mathcal{A}_Q^{(1)} \Bigl|_{D=4 \rm\, div.} = {\beta_0^Q \over \eps} 
{\cal A}^\tree_Q \,.
\label{DivD4YM}
\end{equation}
Here $\beta_0^Q$ is a constant proportional to the one-loop beta
function of the theory.  The only part of the renormalizability of the
theory that we need is that it implies that the color structure of the
divergence must match exactly the color structure of the tree
amplitude.  This holds for any (super) Yang-Mills theory in four
dimensions, though for $\NeqFour$ super-Yang-Mills theory the
beta-function coefficient vanishes, since the theory is ultraviolet
finite~\cite{Mandelstam}.  The color-ordered tree amplitudes satisfy
$U(1)$ decoupling relations,
\begin{equation}
A_Q^\tree (1,2,3,4) + A_Q^\tree (1,3,4,2) + A_Q^\tree (1,4,2,3) = 0 \,,
\label{U1Decoupling}
\end{equation}
which are a simple consequence of the color structure.
Finiteness of the four-point supergravity amplitude follows immediately
by applying \eqns{DivD4YM}{U1Decoupling} to the
supergravity amplitude~(\ref{FourPointSupergravity}), 
\begin{eqnarray}
\mathcal{M}^{(1)}_{Q+16}(1,2,3,4)\Bigr|_{D=4\, \rm{div.}}\! &=&\!
i\bigg(\frac{\kappa}{2}\bigg)^4stA^{\rm tree}_{Q=16}(1,2,3,4)\nn \\
&& \null \hskip .3 cm \times
\bigg[A^{(1)}_{Q}(1,2,3,4) + A^{(1)}_{Q}(1,3,4,2)
 + A^{(1)}_{Q}(1,4,2,3) \bigg]\biggr|_{D=4\, \rm{div.}}\, \nn \\
&=&0\,.
\label{D4OneloopFiniteness}
\end{eqnarray} 

In six dimensions, Yang-Mills theory is not renormalizable. However,
the counterterm has a color structure similar to the $D=4$ one.
For this reason, it is useful to slightly rephrase
the $D=4$ cancellation in terms of a basis of independent color
tensors. As we shall see in the following section, this approach will also
clarify the two-loop finiteness of four-point half-maximal supergravity 
in $D=5$. 

We start with tree level, where there are two independent color tensors
corresponding to the color factors of $s$- and $t$-channel diagrams,
\begin{equation}
b^{(0)}_1  \equiv c^{(0)}_{1234}=\f^{a_1 a_2 b} \f^{b a_3 a_4} \,, \hskip 3 cm 
b^{(0)}_1 \equiv c^{(0)}_{1423} =  \f^{a_2 a_3 b} \f^{b a_4 a_1} \,.
\label{TreeColorTensor}
\end{equation}
The remaining $u$-channel color factor $c^{(0)}_{1324}$ is given
in terms of the previous two by the color Jacobi equation, $c^{(0)}_{1324} =
-b^{(0)}_1 - b^{(0)}_2$.  At one loop there is one additional
independent color tensor (see for example Appendix B of
ref.~\cite{Neq44np}),
\begin{equation}
b^{(1)}_1 \equiv c^{(1)}_{1234}  = \f^{a_1 b_2 b_1} \f^{a_2 b_3 b_2}
 \f^{a_3 b_4 b_3} \f^{a_4 b_1 b_4}\,.
\label{BoxColorTensor}
\end{equation}
The other color factors in the one-loop amplitude
(\ref{OneLoopFourPtGauge}) are given in terms of these color tensors 
after using the color Jacobi identity and the ability to reduce the color
factors with triangle or bubble subdiagrams to tree color tensors. 
For example, we have
\begin{equation}
c^{(1)}_{1342} = b^{(1)}_1 - \frac{1}{2} C_A b^{(0)}_1 \,,
\hskip 1 cm
c^{(1)}_{1423} = b^{(1)}_1 - \frac{1}{2} C_A b^{(0)}_2 \,,
\end{equation}
where $C_A$ is the adjoint representation quadratic Casimir.  For
an $SU(N_c)$ group, $C_A = 2 N_c$ with our nonstandard normalization.

Rewriting the gauge-theory amplitude (\ref{OneLoopFourPtGauge}) in
terms of these independent color tensors gives
\begin{eqnarray}
\nonumber\mathcal{A}^{(1)}_{Q}(1,2,3,4) &=&g^4\bigg[
   b^{(1)}_1 \Bigl(A^{(1)}_{Q}(1,2,3,4)
                 + A^{(1)}_{Q}(1,3,4,2)
                 + A^{(1)}_{Q}(1,4,2,3) \Bigr)\nn \\
 && \hskip 2 cm \null
    - \frac{1}{2} C_A b^{(0)}_1 A^{(1)}_{Q}(1,3,4,2)
    - \frac{1}{2} C_A b^{(0)}_2 A^{(1)}_{Q}(1,4,2,3)     
\biggr]\,. \hskip .3 cm 
\label{OneLoopColorBasis}
\end{eqnarray}
Since the Yang-Mills divergence in $D=4$ contains only the tree color
tensors, it cannot contain the one-loop color tensor $b^{(1)}_1$,
implying that
\begin{equation}
A^{(1)}_Q(1,2,3,4) + A^{(1)}_Q(1,3,4,2) + A^{(1)}_Q (1,4,2,3)
  \Bigr|_{\rm \, div.} = 0\,.
\label{U1DecouplingGen}
\end{equation}
This is equivalent to the tree-level decoupling relation
(\ref{U1Decoupling}), except in \eqn{U1DecouplingGen} there is no explicit
requirement that the divergence of the color-ordered amplitude be
proportional to the tree amplitude, only that the one-loop color
tensor $b^{(1)}_1$ not appear in the divergence.  Thus, we obtain the
vanishing of the supergravity divergence (\ref{D4OneloopFiniteness})
purely from group-theoretic properties of the corresponding gauge theory.

\subsubsection{$D=6$ finiteness}
\label{OneLoopSixDSubSubsection}

\begin{figure}
\begin{center}
\includegraphics[scale=0.5]{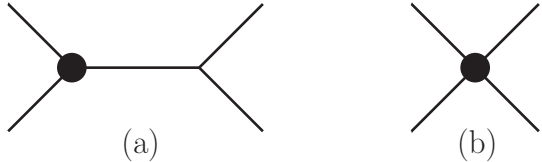}
\caption{The four-point diagrams generated by the $F^3$ counterterm
in pure Yang-Mills at one loop $D=6$ or at two loops in $D=5$.
The large dot indicate an insertion of a counterterm vertex, while a vertex
without a dot represents an ordinary Yang-Mills vertex.  In 
(a) a three-point counterterm vertex appears while in (b) a four-point
counterterm vertex appears.}
\label{CountertermFigure}
\end{center}
\end{figure}
In six dimensions, while the pure graviton $R^3$ counterterm is ruled
out by supersymmetry, naively one might worry about counterterms of the form
$\phi^kR^3$. As we now show, the same group-theoretic cancellations
apply just as well in $D=6$. Since the maximal super-Yang-Mills theory
in six dimensions has $\mathcal{N}=(1,1)$ supersymmetry, the
supergravity theory we are considering is the non-chiral
$\mathcal{N}=(1,1)$ theory, in contrast to the chiral
$\mathcal{N}=(2,0)$ theory.

From simple power-counting considerations, the one-loop $D=6$
pure Yang-Mills counterterm operator is of the form~\cite{F3q},
\begin{equation}
F^3\equiv \f^{abc} F^{a\mu}{}_\nu F^{b\nu}{}_\rho F^{c\rho}{}_\mu\,, 
\label{F3Operator}
\end{equation}
because there are no other gauge-invariant operators of the proper
dimensions that give nonvanishing matrix elements. (The
gauge-invariant operator $D^2F^2$ is also allowed by dimensional
analysis, as noted in ref.~\cite{F3q}. However, these can be removed via
field redefinitions.)  The symmetric color tensor $d^{abc}$ does not
appear because the combination of field strengths has an overall
antisymmetry.  The $F^3$ counterterm is forbidden in super-Yang-Mills
theories because, in a four-dimensional external subspace, it generates
a nonvanishing amplitude with helicities $({\pm}{+}{+}{+})$ that is
disallowed by supersymmetry Ward identities~\cite{SWI}.  However, in
nonsupersymmetric pure Yang-Mills theory in $D=6$, it is a perfectly
valid counterterm with a nonvanishing coefficient.

The key observation is that the counterterm diagrams displayed in
\fig{CountertermFigure} cannot generate color tensors other than the
tree-level ones $b^{(0)}_1$ and $b^{(0)}_2$, defined in
\eqn{TreeColorTensor}.  This follows because the counterterm
three-vertex has a single $\f^{abc}$ and the four-vertex has a pair of
these, so the diagrams in \fig{CountertermFigure} each have a pair of
$\f^{abc}$'s.  Since the one-loop color tensor $b^{(1)}_1$ is built
from four $\f^{abc}$'s, it cannot appear in the Yang-Mills divergence.
Thus, the situation is quite similar to the $D=4$ case where only tree
color tensors can appear in the divergence.

Following the $D=4$ discussion, we demand that the one-loop color tensor
$b^{(1)}_1$ not appear in the divergence.
From \eqn{OneLoopColorBasis} we see that the $U(1)$ decoupling equation
(\ref{U1DecouplingGen}) holds for the $D=6$ divergences.
Plugging this into \eqn{FourPointSupergravity}
immediately shows that the one-loop divergence for $D=6$ pure
supergravity with 16 or more supercharges must vanish:
\begin{equation}
\mathcal{M}^{(1)}_{Q+ 16}(1,2,3,4) \biggr|_{D=6\, \rm div.} = 0\,.
\end{equation}
For $Q>0$, the divergence not only vanishes because the decoupling equation
(\ref{U1DecouplingGen}) holds, but also because $F^3$ is not a valid
supersymmetric counterterm of the corresponding gauge theory.

It is straightforward to confirm that the decoupling identity
(\ref{U1DecouplingGen}) holds using the explicit forms of pure
Yang-Mills counterterm amplitudes generated by the diagrams in
\fig{CountertermFigure}.  For example, the all-plus helicity
counterterm amplitude in a four-dimensional subspace
is~\cite{DixonF3}
\begin{equation}
A(1^+,2^+,3^+,4^+)= \frac{\alpha}{\eps}
 \frac{stu}
   {\langle12\rangle\langle23\rangle\langle34\rangle\langle41\rangle}\,,
\end{equation}
where $\alpha$ is a proportionality constant which can be fixed by
explicit computation, but its value is unimportant for our discussion.
This expression does indeed satisfy the required $U(1)$ decoupling
identity (\ref{U1DecouplingGen}) because
\begin{equation}
\frac{1}{\langle12\rangle\langle23\rangle\langle34\rangle\langle41\rangle}+
\frac{1}{\langle13\rangle\langle34\rangle\langle42\rangle\langle21\rangle}+
\frac{1}{\langle14\rangle\langle42\rangle\langle23\rangle\langle31\rangle}=0\,.
\end{equation} 

The vanishing of the $Q = 16$ counterterm can also be understood using
a color-trace basis. Using the double-copy formula,
ref.~\cite{Schnitzer} showed that the one-loop $Q\ge 16$ supergravity
amplitudes can be written as the double-trace Yang-Mills amplitude
multiplied by a kinematic-dependent factor. This immediately leads us
to conclude that six-dimensional $\mathcal{N}\geq4$ supergravity must
be one-loop ultraviolet finite since counterterm amplitudes generated
with the $F^3$ operator do not contain a double-trace contribution.

We have also computed the coefficient of the $D=6$ divergence of
half-maximal supergravity using the procedure of ref.~\cite{N4grav}
and have confirmed that it vanishes.  In this construction, one copy is
the maximally supersymmetric Yang-Mills amplitude, while the second copy is
based on ordinary Feynman rules.  As mentioned earlier, the pure
Yang-Mills numerators do not need to satisfy the duality
\eqn{BCJDuality} since the maximal super-Yang-Mills side already
does.   A key simplifying feature of
this method is that pure Yang-Mills numerators are required only for
the box diagram since four-point maximal super-Yang-Mills
numerators vanish for all other diagram topologies.  In addition,
there are no subdivergences since we are dealing here with one
loop.  We find the divergence cancels completely, in complete
agreement with the above much simpler counterterm considerations.

\subsubsection{$D=8$ divergences}

We now consider the $D=8$ case.  From ref.~\cite{F3q}, the pure
nonsupersymmetric Yang-Mills divergence is described by an $F^4$
operator of the form,
\begin{eqnarray}
&&F^4 = c^{a b c d}
 \biggl[ d_1 F^{a\,\mu \nu} F^{b}{}_{\nu\sigma}F^{c\,\sigma\rho} F^{d}{}_{\rho\mu}
+ d_2 F^{a\,\mu \nu} F^{b}{}_{\nu\sigma} F^{c}{}_{\rho\mu} F^{d\,\sigma\rho} \nn \\
&&\hskip 1.8 cm + d_3 F^{a\,\mu\nu} F^{b}{}_{\nu\mu} F^{c\,\rho\sigma} F^{d}{}_{\sigma\rho}
+ d_4 F^{a\,\mu\nu} F^{b\,\rho\sigma} F^{c}{}_{\nu\mu} F^{d}{}_{\sigma\rho}
 \biggr] \,,
\label{YMF4Operator}
\end{eqnarray}
where 
\begin{equation}
c^{abcd} \equiv 
\f^{a\,e_1 e_2} \f^{b\,e_2 e_3} \f^{c\,e_3 e_4} \f^{d\,e_4 e_1} \,,
\label{OneloopColor}
\end{equation}
is the box-diagram color factor using the normalization in
\eqn{fabcdef}. Only the linearized part of the field strength
contributes to the divergent part of the four-point amplitude,
$F_{\mu\nu}\equiv\partial_{[\mu} A_{\nu]}$.  In $D=8$ at one loop the
constants appearing in the operator are
\begin{eqnarray}
&&d_1 = \frac{g^4}{8} \frac{1} {(4 \pi)^4} \frac{1}{\eps} \, 
  \frac{670+D_s}{2520}\,, \hskip 2 cm
d_2 = \frac{g^4}{8} \frac{1} {(4 \pi)^4} \frac{1}{\eps} \, 
   \frac{166+D_s}{420}, \nn \\
&&d_3 = \frac{g^4}{8} \frac{1} {(4 \pi)^4} \frac{1}{\eps} \, 
  \frac{17D_s-706}{5040}\,, \hskip 1.575 cm
d_4 = \frac{g^4}{8} \frac{1} {(4 \pi)^4} \frac{1}{\eps} \, 
   \frac{D_s-338}{10080} \,,
\end{eqnarray}
where $D_s$ is a state-counting parameter.  It comes from contracting
the metric $\eta_{\mu\nu}$ from gluon propagators around the loop.  In
pure Yang-Mills theory in eight dimensions, we take the state-counting
parameter to be $D_s = 8$.  (This is equivalent to the
four-dimensional helicity regularization scheme~\cite{FDH} but with
the state count adjusted to match the one of eight dimensions.) The
divergence was first derived with $D_s=8$ in ref.~\cite{vdV}.
For the four-gluon amplitude at one loop, $n_s=D_s-8$ counts additional
minimally-coupled scalars circulating in the loop.  The key difference
between the $D=8$ case and the previous $D=4,6$ cases is that the
gauge-theory divergence contains the independent color tensor
$b^{(1)}$. Thus the $U(1)$ decoupling equation (\ref{U1DecouplingGen})
does {\it not} hold.

The amplitude is given by replacing the vector potential with the
polarization vector $\pol_j$, giving a polarization field strength for
each leg $j$,
\begin{equation}
F_j^{\mu\nu} \equiv i( k_j^\mu \pol_j^\nu - k_j^\nu \pol_j^\mu) \,.
\label{FtoAmp}
\end{equation}
For notational convenience, we define the contractions of these
polarization field strengths as
\begin{equation}
(F_i F_j) \equiv F_i^{\mu\nu} F_{j\mu\nu},\hskip 1 cm 
(F_{i}F_{j}F_{k}F_{l})\equiv F_i\,^{\mu\nu}F_{j\nu\rho}F_k\,^{\rho\sigma}F_{l\sigma\mu}\,.
 \end{equation}
In terms of these, the nonvanishing divergence in the
nonsupersymmetric pure Yang-Mills amplitude is

\begin{eqnarray}
&&\hskip -.4 cm 
 \mathcal{A}^{(1)}_{Q=0}(1,2,3,4) \Bigr|_{D=8\, \rm div.} \hskip -.2 cm 
 \nn \\
&&\hskip .4 cm = \frac{i}{8 \eps } \frac{1} {(4 \pi)^4}   \, 
g^4 c^{a_1 a_2 a_3 a_4}  \label{YMD8Div}
\Biggl[ 8\frac{670+D_s}{2520} (F_1F_2F_3F_4)+4\frac{166+D_s}{420}\Big((F_1F_3F_4F_2)+(F_1F_4F_2F_3)\Big)
  \nn \\
&&\hskip 2.2 cm \null
+4\,\frac{17D_s-706}{5040} \Big((F_1 F_2) (F_3 F_4) 
+ (F_2 F_3) (F_4 F_1)\Big)+8\frac{D_s-338}{10080}(F_1 F_3) (F_2 F_4)\Biggr] \nn \\
&&\hskip 1.3 cm \null
+\mathrm{cyclic}(2,3,4)\,,
\hskip 1 cm \
\end{eqnarray}
where, as before, `cyclic$(2,3,4)$' indicates that one should
include the two cyclic permutations of legs 2, 3 and 4
along with their color indices.  Matching \eqn{YMD8Div} with
\eqn{OneLoopFourPtGauge} and replacing color factors by the
corresponding $Q=16$ super-Yang-Mills numerators immediately gives the
explicit form of the $Q=16$ eight-dimensional supergravity divergence:
\begin{eqnarray}
\mathcal{M}^{(1)}_{Q=16}(1,2,3,4)\bigg|_{D=8\, \rm div.} \hskip -.2 cm &=&
-\frac{1}{\eps}  \frac{1} {(4 \pi)^4}  \bigg(\frac{\kappa}{2}\bigg)^4stA^{\rm tree}_{Q = 16}(1,2,3,4) \label{GravityOneLoopD8}\\
&\times& \hskip -.2 cm 
\bigg[ \frac{238+D_s}{360}(F_1F_2F_3F_4) + \frac{D_s-50}{288}(F_1 F_2)(F_3 F_4)
\bigg] +{\rm cyclic(2,3,4)} \,,\nn 
\end{eqnarray} 
where $D_s = 8$ in the pure supergravity case.
The factor $A^{\rm
  tree}_{Q= 16}(1,2,3,4)$ is just the maximally supersymmetric
four-point tree amplitude, for any of the states in the theory.  
The corresponding
states in the $Q=16$ supergravity theory are just the tensor product
of these states with gluon states of the pure nonsupersymmetric Yang-Mills
theory.

The explicit four-graviton $R^4$ counterterm for half-maximal
supergravity in $D=8$ is given in ref.~\cite{CountertermsDunbar}.  It
is built from the seven linearly independent $R^4$ forms in $D=8$
\cite{RiemannTensor} (in $D<8$ these are no longer independent):
\begin{eqnarray}
T_1&=&(R_{\mu\nu\rho\sigma}R^{\mu\nu\rho\sigma})^2, \nn \\
T_2&=&R_{\mu\nu\rho\sigma}R^{\mu\nu\rho}_{\hphantom{\mu\nu\rho}\lambda}R_{\gamma\delta\kappa}^{\hphantom{\gamma\delta\kappa}\sigma}R^{\gamma\delta\kappa\lambda}, \nn \\
T_3&=&R_{\mu\nu\rho\sigma}R^{\mu\nu}_{\hphantom{\mu\nu}\lambda\gamma}R^{\lambda\gamma}_{\hphantom{\lambda\gamma}\delta\kappa}R^{\rho\sigma\delta\kappa}, \nn \\
T_4&=&R_{\mu\nu\rho\sigma}R^{\mu\nu}_{\hphantom{\mu\nu}\lambda\gamma}R^{\rho\lambda}_{\hphantom{\rho\lambda}\delta\kappa}R^{\sigma\gamma\delta\kappa}, \nn \\
T_5&=&R_{\mu\nu\rho\sigma}R^{\mu\nu}_{\hphantom{\mu\nu}\lambda\gamma}R^{\rho\hphantom{\delta}\lambda}_{\hphantom{\rho}\delta\hphantom{\lambda}\kappa}R^{\sigma\delta\gamma\kappa}, \nn \\
T_6&=&R_{\mu\nu\rho\sigma}R^{\mu\hphantom{\lambda}\rho}_{\hphantom{\mu}\lambda\hphantom{\rho}\gamma}R^{\lambda\hphantom{\delta}\gamma}_{\hphantom{\lambda}\delta\hphantom{\gamma}\kappa}R^{\nu\delta\sigma\kappa}, \nn \\
T_7&=&R_{\mu\nu\rho\sigma}R^{\mu\hphantom{\lambda}\rho}_{\hphantom{\mu}\lambda\hphantom{\rho}\gamma}R^{\lambda\hphantom{\delta}\nu}_{\hphantom{\lambda}\delta\hphantom{\nu}\kappa}R^{\gamma\delta\sigma\kappa} \,.
\end{eqnarray}
On shell the combination,
\begin{equation}
-\frac{T_1}{16}+T_2-\frac{T_3}{8}-T_4+2T_5-T_6+2T_7\,,
\label{onShellVanish}
\end{equation}
is a total derivative, so only 6 of the $T_i$ are independent on shell.  This gives us some freedom in how we write the explicit counterterm, which we give as \cite{CountertermsDunbar}
\begin{eqnarray}
&&\frac{1}{\epsilon}\frac{1}{(4\pi)^4}\frac{1}{11520}
[(-126+3D_s)T_1+(1968-24D_s)T_2+(-252+6D_s)T_3 \nn \\
&&\hskip 2.6cm 
 \null +(8-4D_s)T_4+3840T_5-1920T_6+(-3776-32D_s)T_7 ] \,,
\label{DunbarR4}
\end{eqnarray}
where there is a relative $i$ between the operators and amplitudes.  The
appropriate powers of the coupling are generated by expanding the metric
around flat space, $g_{\mu\nu} = \eta_{\mu\nu}+\kappa h_{\mu\nu}$.

Since the kinematic numerators for one-loop four-point $\mathcal{N}=4$
super-Yang-Mills theory are independent of loop momenta, we can also
write the counterterm in a manner more suggestive of the double-copy
structure.  The four-point numerators in \eqn{FourPointNumerators} are
given by an $F^4$ operator:
\begin{equation}
F^4=-2\left[(F_1F_2F_3F_4)-\frac{1}{4}(F_1F_2)(F_3F_4)+\mathrm{cyclic}(2,3,4)\right]\,.
\end{equation}
Up to an overall constant including color factors, this is the same $F^4$
counterterm for four-point one-loop $\mathcal{N}=4$ sYM in $D=8$.
Using this operator as a replacement for the
kinematic numerator in \eqn{GravityOneLoopD8}, we build an $R^4$
counterterm by making the association,
\begin{equation}
F_{i\,\mu\nu}F_{i\,\rho\sigma}\rightarrow -2R_{i\,\mu\nu\rho\sigma}\,.
\label{FFtoR}
\end{equation}
At the linearized level, both terms in \eqn{FFtoR} give the same
contribution to the amplitude.  On the gravity side, we replace the graviton field $h_{\mu\nu}$ by the polarization tensor $\epsilon_{\mu\nu}$, which can itself be replaced by the symmetrization of two polarization vectors $\epsilon_{\mu\nu}\rightarrow\epsilon_{(\mu}\epsilon_{\nu)}$.  For the case of gravitons, we treat the two polarization vectors as being identical since the two possible replacements are $\epsilon_{\mu\nu}^{++}\rightarrow\epsilon_{\mu}^+\epsilon_{\nu}^+$ and $\epsilon_{\mu\nu}^{--}\rightarrow\epsilon_{\mu}^-\epsilon_{\nu}^-$.  We then have
\begin{eqnarray}
R_{\mu\nu\rho\sigma}&=&\eta_{\mu\lambda}\left(\partial_{\rho}\Gamma^{\lambda}_{\nu\sigma}-\partial_{\sigma}\Gamma^{\lambda}_{\nu\rho}\right) \nn \\
&=&\frac{1}{2}ik_{\rho}(ik_{\sigma}\epsilon_{\mu}\epsilon_{\nu}+ik_{\nu}\epsilon_{\mu}\epsilon_{\sigma}-ik_{\mu}\epsilon_{\nu}\epsilon_{\sigma})-\frac{1}{2}ik_{\sigma}(ik_{\rho}\epsilon_{\mu}\epsilon_{\nu}+ik_{\nu}\epsilon_{\mu}\epsilon_{\rho}-ik_{\mu}\epsilon_{\nu}\epsilon_{\rho}) \nn \\
&=&\frac{1}{2}(k_{\mu}\epsilon_{\nu}-k_{\nu}\epsilon_{\mu})(k_{\rho}\epsilon_{\sigma}-k_{\sigma}\epsilon_{\rho})\,.
\end{eqnarray}
Comparing to the polarization field strength tensor in \eqn{FtoAmp}
gives us the replacement rule (\ref{FFtoR}). 
After taking into account permutations, this replacement rule gives us the following contributing $R^4$ forms:
\begin{eqnarray}
U_1&=&R_{\mu\nu\lambda\gamma}R^{\nu\hphantom{\rho}\gamma}_{\hphantom{\nu}\rho\hphantom{\gamma}\delta}R^{\rho\hphantom{\sigma}\delta}_{\hphantom{\rho}\sigma\hphantom{\delta}\kappa}R^{\sigma\mu\kappa\lambda}, \nn \\
U_2&=&R_{\mu\nu\lambda\gamma}R^{\nu}_{\hphantom{\nu}\rho\delta\kappa}R^{\rho\hphantom{\sigma}\gamma\delta}_{\hphantom{\rho}\sigma}R^{\sigma\mu\kappa\lambda}, \nn \\
U_3&=& (R_{\mu\nu\rho\sigma}R^{\mu\nu\rho\sigma})^2, \nn \\
U_4&=&R_{\mu\nu\lambda\gamma}R^{\mu\nu}_{\hphantom{\mu\nu}\delta\kappa}R_{\rho\sigma}^{\hphantom{\rho\sigma}\lambda\gamma}R^{\rho\sigma\delta\kappa}, \nn \\
U_5&=&R_{\mu\nu\lambda\gamma}R^{\nu\hphantom{\rho}\lambda\gamma}_{\hphantom{\nu}\rho}R^{\rho}_{\hphantom{\rho}\sigma\delta\kappa}R^{\sigma\mu\delta\kappa}, \nn \\
U_6&=&R_{\mu\nu\lambda\gamma}R^{\nu}_{\hphantom{\nu}\rho\delta\kappa}R^{\rho\hphantom{\sigma}\lambda\gamma}_{\hphantom{\rho}\sigma}R^{\sigma\mu\delta\kappa} \,,
\end{eqnarray}
and the counterterm is given by
\begin{eqnarray}
&&\frac{1}{\epsilon}\frac{1}{(4\pi)^4}\frac{1}{23040}\left[(-3808-16D_s)U_1+(-7616-32D_s)U_2+(-250+5D_s)U_3\right. \nn \\
&&\hskip 2.4cm \left.+(-500+10D_s)U_4+(3904-32D_s)U_5+(1952-16D_s)U_6\right]\,.
\label{DoubleCopyR4}
\end{eqnarray}
At the linearized level this is equivalent to \eqn{DunbarR4}, but
instead the index structure has been reorganized to expose the
double-copy structure of gravity. 

In terms of spinor helicity in a four-dimensional external subspace
for the four-graviton case with external helicities
$(1^{+},2^{+},3^{-},4^{-})$, the divergence in $D=8$ is
\begin{eqnarray}
\mathcal{M}^{(1)}_{Q=16}(1^{+},2^{+},3^{-},4^{-})\bigg|_{D=8\ \rm div.}
=\frac{i}{\eps}  \frac{1} {(4 \pi)^4} 
\bigg(\frac{\kappa}{2}\bigg)^4\,\frac{58+D_s}{180} \langle 3 4 \rangle^4 [12]^4\,,
\end{eqnarray}
where we have plugged in spinor helicity for the
$(1^{+},2^{+},3^{-},4^{-})$ configuration on the right side of
\eqn{GravityOneLoopD8}.  Similarly, any of the other helicity
amplitudes can be extracted from \eqn{GravityOneLoopD8}.

As in $D=4,6$ dimensions, we have also used the procedure described in
ref.~\cite{N4grav} for explicitly computing the divergences in 
half-maximal supergravity, finding agreement with the divergence in
\eqn{GravityOneLoopD8}.

\subsection{Five points at one loop}
\label{FivePointOneloopSubsection}

To make the vanishing of ultraviolet divergences of half-maximal
supergravity in $D=4$ and $D=6$ manifest at five-point one-loop, we write the
one-loop supergravity amplitude (\ref{SimpleOneloopGravitySusy}) in terms
of a basis of six independent $\beta$'s (defined in
\eqn{OneLoopPentagonNumerators}):
\begin{eqnarray}
\nonumber&&{\cal M}_{Q+16}^\oneloop (1,2, 3,4,5)=i\left(\frac{\kappa}{2}\right)^5 \\
\nonumber&&\hspace{.9cm}
\times\left(\beta_{12345}(A^{(1)}_Q(1,2,3,4,5)+A^{(1)}_Q(2,1,3,4,5)+A^{(1)}_Q(2,3,1,4,5)
             +A^{(1)}_Q(2,3,4,1,5))\right.\nn \\
&&\hspace{1.1 cm} \null 
+\beta_{12354}(A^{(1)}_Q(3,1,2,5,4)+A^{(1)}_Q(1,3,2,5,4)+A^{(1)}_Q(1,2,3,5,4)
             +A^{(1)}_Q(1,2,5,3,4))\nn \\
\nn&&\hspace{1.1cm}\null
+\beta_{12435}(A^{(1)}_Q(2,1,4,3,5)+A^{(1)}_Q(1,2,4,3,5)+A^{(1)}_Q(1,4,2,3,5)
             +A^{(1)}_Q(1,4,3,2,5))\nn \\
\nn&&\hspace{1.1cm}\null
+\beta_{12453}(A^{(1)}_Q(4,1,2,5,3)+A^{(1)}_Q(1,4,2,5,3)+A^{(1)}_Q(1,2,4,5,3)
             +A^{(1)}_Q(1,2,5,4,3))\nn \\
&&\hspace{1.1cm}\null
+\beta_{13245}(A^{(1)}_Q(5,1,3,2,4)+A^{(1)}_Q(1,5,3,2,4)+A^{(1)}_Q(1,3,5,2,4)
             +A^{(1)}_Q(1,3,2,5,4))\nn \\
&&\hspace{1.1cm}\null
\left.+\beta_{13425}(A^{(1)}_Q(3,1,4,2,5)+A^{(1)}_Q(1,3,4,2,5)+A^{(1)}_Q(1,4,3,2,5)
             +A^{(1)}_Q(1,4,2,3,5))\right)\,. \nn\\
{}
\label{FivePtInBasis}
\end{eqnarray}
This expression is valid for all amplitudes where the external
gluons on the super-Yang-Mills side of the double copy are in an MHV
configuration in a four-dimensional subspace.  The \MHVbar{} result is
just the parity conjugate.  From the form (\ref{FivePtInBasis}), it is
clear that when the gauge-theory divergences satisfy five-point $U(1)$
decoupling relations,
\begin{equation} 
A^{(1)}_Q(1,2,3,4,5)+A^{(1)}_Q(2,1,3,4,5)
+A^{(1)}_Q(2,3,1,4,5)+A_Q^{(1)}(2,3,4,1,5) 
  \Bigr|_{\rm div.} = 0 \,,
\label{FivePoinDecoupling}
\end{equation}
and relabelings thereof, the supergravity amplitude
(\ref{FivePtInBasis}) is finite, in much the same way as at four
points. At tree level, these decoupling identities and their related Kleiss-Kuijf
relations~\cite{KleissKuijf} are purely a consequence of color
considerations~\cite{DixonMaltoniColor}. Alternatively, as noted in
ref.~\cite{BCJ}, they follow from the requirement that the
color-ordered amplitudes can be described by diagrams with
antisymmetric cubic vertices.  As discussed above for the four-point
case, in both $D=4$ and $D=6$ the Yang-Mills counterterms 
generate exactly the same color structures as at tree level, 
 so the decoupling equation (\ref{FivePoinDecoupling}) indeed holds.  Therefore, we
immediately conclude that
\begin{equation}
\mathcal{M}^{(1)}_{Q+16}(1,2,3,4,5) \Bigr|_{D=4\, \rm div.} = 0  \,, \hskip 2 cm 
\mathcal{M}^{(1)}_{Q+16}(1,2,3,4,5) \Bigr|_{D=6\, \rm div.} = 0  \,.
\end{equation}
Had we used a different basis of $\beta$'s, there could have been more
terms multiplying a given $\beta$, but at the end the divergences still
cancel due to the $U(1)$ decoupling identity.

We have also directly confirmed the vanishing of the divergences in
$D=4,6$, and computed the nonvanishing divergence of half-maximal
supergravity in $D=8$, using the procedure in ref.~\cite{N4grav}.  
In this procedure we take one
copy to be maximal $Q = 16$ super-Yang-Mills theory and the other copy
pure nonsupersymmetric Yang-Mills theory.  From the double-copy
formula (\ref{SimpleOneloopGravitySusy}), we have
\begin{equation}
\mathcal{M}^{(1)}_{Q=16}(1,2,3,4,5)=-\left(\frac{\kappa}{2}\right)^5
\sum_{{\cal S}_5}\left(\frac{1}{10}\beta_{12345}
\int \frac{d^{D}p} {(2\pi)^D} \frac{n_{12345}}{\prod_{\alpha_j}{p^2_{\alpha_j}}}
+\frac{1}{4}\gamma_{12}\int \frac{d^{D}p} {(2 \pi)^D}
 \frac{n_{[12]345}}{\prod_{\alpha_j}{p^2_{\alpha_j}}}\right)\,.
\label{SugraOneLoopFivePt}
\end{equation}
Here $n_{12345}$ and $n_{[12]345}$ are numerators of pure Yang-Mills
pentagon (shown in \fig{OneLoopDiagramsFigure}(b)) and box diagrams
(shown in \fig{OneLoopDiagramsFigure}(c)) respectively, derived from
Feynman diagrams in Feynman gauge. As described in ref.~\cite{N4grav},
the derived numerators include ghost contributions and contributions
from four-point contact terms assigned according to their color
factors. The $\beta_{12345}$ given in \eqn{OneLoopPentagonNumerators}
and $\gamma_{12}$ given in \eqn{gammas} are the corresponding pentagon
and box numerators of maximal super-Yang-Mills theory.  The
propagators are those of each graph.  The sum ${\cal S}_5$ runs over
all $5!$ permutations of the external legs, with symmetry factors
included to adjust for the overcount. The symmetry factors for
\fig{OneLoopDiagramsFigure}(b) and \fig{OneLoopDiagramsFigure}(c) are
$10$ and $4$ respectively.  The expression (\ref{SugraOneLoopFivePt})
is valid when the external gluons on the super-Yang-Mills side of the
double copy are in an MHV configuration in the four-dimensional
external subspace.  The \MHVbar{} configuration is obtained using
parity.

Restricting the integrals to the divergent part, we find the
divergences in $D=4,6$ to vanish, as was the case at four points.  In
$D=8$ we find a nonvanishing divergence, the explicit form of which we
have included in an accompanying Mathematica
attachment~\cite{AttachedFile}.  The first two terms of this
expression are
\begin{eqnarray}
\mathcal{M}^{(1)}_{Q=16}(1,2,3,4,5)\Bigr|_{D=8\, \rm div.} &=&
\frac{1}{(4\pi)^4}\biggl(\frac{\kappa}{2}\biggr)^5 \Biggl[
\frac{238+D_s}{180\sqrt{2}\epsilon}
\gamma_{34}\,\pol_1 \cdot \pol_4 k_1\cdot \pol_2 k_1 \cdot \pol_5 
        k_2 \cdot \pol_3  \\
&&\null \hskip 1.2 cm 
-\frac{D_s-122}{180\sqrt{2}\epsilon} \gamma_{14} \frac{\pol_1 \cdot\pol_2
 \pol_4 \cdot \pol_5  k_1 \cdot \pol_3 s_{23} s_{24}}{ s_{45}}+\cdots
\Biggl]\,, \nn
\label{fivePtDEq8}
\end{eqnarray}
where the $\gamma_{ij}$ are the box numerators defined in \eqn{gammas}
and the $\pol_i$ are gluon polarization vectors.  As always the
supergravity states are simply tensor products of the maximal
super-Yang-Mills states with those of pure Yang-Mills theory.

\begin{figure}
\begin{center}
\includegraphics[scale=0.5]{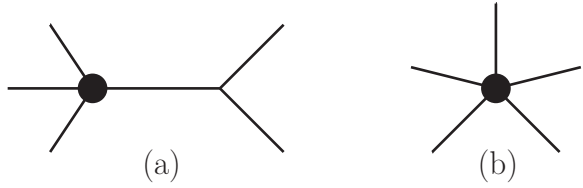}
\caption{The counterterm diagrams describing the one-loop divergences
  of either pure Yang-Mills theory or half-maximal supergravity in
  $D=8$.  The large dots indicate an insertion
  of a counterterm vertex generated by either an $F^4$ operator in Yang-Mills
  theory or an $R^4$ operator in supergravity.}
\label{Counterterm5Figure}
\end{center}
\end{figure}

As a nontrivial check, we have reproduced
the $D=8$ result in an additional way, which we briefly summarize.
We used the Yang-Mills $F^4$ operator in \eqn{YMF4Operator} to
obtain the five-point pure Yang-Mills divergence using the Feynman
diagrams illustrated in \fig{Counterterm5Figure}.  Plugging the
color-ordered Yang-Mills divergences into \eqn{SimpleOneloopGravitySusy}
yields the gravity divergence:
\begin{equation}
\mathcal{M}^{(1)}_{Q=16}(1,2,3,4,5)\Bigr|_{D=8\, \rm div.}=
i\biggl( \frac{\kappa}{2} \biggr)^5
\sum_{{\cal S}_5/({\cal Z}_5\times {\cal Z}_2)}\, \beta_{12345} \, 
A_{Q=0}^{(1)}(1,2,3,4,5)\Bigr|_{D=8\, \rm div.}\,.
\end{equation}
Remarkably, this suggests that the entire five-loop divergence in
$D=8$ for half-maximal supergravity is contained in the operators that
describe four-point divergences and that no further independent
operators should appear at five points. 

As a first test of this, we used the $R^4$ counterterm as determined
at four points to compute the five-graviton divergence, again using
diagrams of the form shown in \fig{Counterterm5Figure}.  We used both
forms of the counterterm (\eqn{DunbarR4} and \eqn{DoubleCopyR4}); in
both cases we find agreement with \eqn{fivePtDEq8} for
four-dimensional external graviton states.  For cases with other
external states, we suspect again all two-loop supergravity divergences are
locked to the four-point divergences given that no new independent
five- or higher-point counterterms arise in pure Yang-Mills theory (by simple
gauge invariance and dimensional considerations).  It would be
interesting to investigate this further.

\subsection{Comments on the four-point one-loop $\mathcal{N}=4$
 gravity-matter system}

The above group theoretic analysis can also be applied to understand
the divergence structure of $Q \ge 16$ supergravity with matter.  A
particularly interesting case is $\mathcal{N}=4$ supergravity in $D=4$
coupled to $n_v$ $\mathcal{N}=4$ vector multiplets.  These theories
naturally arise from dimensional reduction of half-maximal pure
supergravity models in higher dimensions.  Over 30 years ago, Fischler
showed that this theory is ultraviolet divergent at one
loop~\cite{Fischler}.  This result can be simply understood from the
double-copy vantage point.

In the double-copy picture, $\NeqFour$ supergravity amplitudes with
vector multiplets are constructed using $\mathcal{N}=4$
super-Yang-Mills amplitudes for one copy and a Yang-Mills theory with
adjoint scalars that interact with gluons.  In the latter theory, the
only allowed interactions of the scalar are the standard minimal
interactions with gluons or self interactions via a $\phi^4$ operator
for the second copy.  With either interaction, simple
renormalizability constraints in $D=4$ show that the only gauge-theory
operators that can act as counterterms are the form $F^2$, $(D_\mu
\phi)^2$ or $\phi^4$.  The first two operators generate amplitudes
containing only tree-level color tensors, so the divergences satisfy
$U(1)$ (\ref{U1DecouplingGen}) decoupling relations.  Hence from
\eqn{FourPointSupergravity}, we immediately have that amplitudes with
only supergravity multiplet states on the external lines or with
two-graviton and two-vector multiplet states are finite irrespective
of the number of vector multiplets.  However, one-loop four-point
amplitudes where all external legs belong to the matter multiplet are
different. In the scalar-Yang-Mills system, the four-scalar
counterterm operator with a one-loop color tensor of the form,
\begin{equation}
 c^{abcd} \phi^{a} \phi^{b} \phi^{c} \phi^{d} \,,
\end{equation}
is allowed, where $c^{abcd}$ is defined in \eqn{OneloopColor}. 
Here the generated divergence does not satisfy $U(1)$ decoupling,
and when fed through \eqn{FourPointSupergravity}, 
the corresponding supergravity amplitude diverges.
 Indeed this is consistent with the divergence in the
four-matter-multiplet amplitude found long ago by
Fischler~\cite{Fischler}. The same conclusion was also reached in
ref.~\cite{Arkady} with a corrected overall constant.

The case of $D=6$ is a bit different.  Here a divergence for the
two-matter two-gravity matrix element appears.  The presence of this
supergravity divergence can be understood from the double-copy
viewpoint as originating from a counterterm of the nonsupersymmetric
scalar-Yang-Mills system:
\begin{equation}
c^{abcd} F^{a}{}_{\!\mu\nu} 
F^{b}{}^{\mu\nu} \phi^{c} \phi^{d} \,. 
\end{equation}
In the double-copy formula, when this is combined with maximally
supersymmetric Yang- Mills theory, we obtain a nonvanishing
two-graviphoton and two-matter-photon counterterm of the form
$D^2F^4$. This is related by supersymmetry to the two-graviton
two-matter-photon counterterm $R^2F^2$. While we have not explicitly
computed this divergence, it would be an interesting exercise to do
so.

\section{Half-maximal supergravity at two loops}
\label{TwoloopUVSection}

We now turn to the main topic of this paper, which is the divergence
structure of half-maximal supergravity at two loops. We follow similar
reasoning as for the cases of $D=4,6$ at one loop.  In particular, we
demonstrate that the same cancellations that prevent forbidden color
structures from appearing in pure Yang-Mills divergences are
responsible for making the half-maximal pure supergravity two-loop
four-point amplitude finite in $D=5$.  On dimensional grounds, we
expect the $D=5$ two-loop four-point counterterm of supergravity to be
a supersymmetric completion of an $R^4$
operator~\cite{GrisaruSiegel,BossardHoweStelle5D}.  Nevertheless the
corresponding divergence vanishes.  We also explicitly demonstrate the
ultraviolet finiteness of a subset of five-point amplitudes with
external states in a four-dimensional subspace; specifically we look
at those amplitudes where the external supergravity states are those
obtained as a tensor product of gluon states in the four-dimensional
subspace.  Besides explaining the lack of a two-loop divergence in
these amplitudes in $D=5$, we also obtain the explicit value of the
four-point divergence in $D=6$.

\subsection{Four-point divergence cancellations at two loops}

\subsubsection{Group theory considerations}

Ordinary nonsupersymmetric Yang-Mills theory in $D=5$ is, of course,
divergent.  At two loops in $D=5$, the available counterterm in this
theory is of the same $F^3$ form (\ref{F3Operator}) as at one loop in
$D=6$.  In $D=5$ there are no one-loop divergences in dimensional
regularization, so we do not need to concern ourselves with
subdivergences.

Following the same logic as at one loop, we impose the constraint that
the $F^3$ operator generates only the tree-level color
structures. Using the color basis described in Appendix B of
ref.~\cite{Neq44np} (see also ref.~\cite{Naculich}), we express the
color factors in \eqn{YMTwoloop} in terms of the independent tree and
one-loop color tensors given in
\eqns{TreeColorTensor}{BoxColorTensor}, as well as two additional
two-loop color tensors, $b^{(2)}_1$ and $b^{(2)}_2$.  For the planar
color factors we have
\begin{eqnarray}
c^{\rm P}_{1234} &=& b^{(2)}_1 \,, \hskip 3.9 cm 
c^{\rm P}_{2341} = b^{(2)}_2 \,, \nn \\
c^{\rm P}_{3421} &=& b^{(2)}_1 - \frac{1}{4} C_A^2 b^{(0)}_1 \,, \hskip 2 cm 
c^{\rm P}_{1423} = b^{(2)}_2 - \frac{1}{4} C_A^2 b^{(0)}_2 \,, \nn \\
c^{\rm P}_{1342} &=& -b^{(2)}_1 - b^{(2)}_2 + \frac{3}{2} C_A b^{(1)}_1 
                      - \frac{1}{4} C_A^2 b^{(0)}_1 \,, \nn \\
c^{\rm P}_{4231} &=& -b^{(2)}_1 - b^{(2)}_2 + \frac{3}{2} C_A b^{(1)}_1 
                      - \frac{1}{4} C_A^2 b^{(0)}_2\,.
\end{eqnarray}
Similarly for the nonplanar color factors we have
\begin{eqnarray}
c^{\rm NP}_{1234} &=& c^{\rm P}_{1 2 3 4} - \frac{1}{2} C_A b^{(1)}_1\,,\nn \\
c^{\rm NP}_{2341} &=& c^{\rm P}_{2 3 4 1} - \frac{1}{2} C_A b^{(1)}_1 \,, \nn\\
c^{\rm NP}_{3421} &=& c^{\rm P}_{3 4 2 1} - \frac{1}{2} C_A b^{(1)}_1 
                           + \frac{1}{4} C_A^2 b^{(0)}_1 \,, \nn \\
c^{\rm NP}_{1423} &=& c^{\rm P}_{1423} - \frac{1}{2} C_A b^{(1)}_1 
                           + \frac{1}{4} C_A^2 b^{(0)}_2  \,, \nn \\
c^{\rm NP}_{1342} &=& c^{\rm P}_{1342} - \frac{1}{2} C_A b^{(1)}_1 
                           + \frac{1}{4} C_A^2 b^{(0)}_1 \,, \nn \\
c^{\rm NP}_{4231} &=&  c^{\rm P}_{4231} - \frac{1}{2}  C_A b^{(1)}_1 
                           + \frac{1}{4} C_A^2 b^{(0)}_2 \,.
\end{eqnarray}
Inserting these into the gauge-theory amplitude (\ref{YMTwoloop}) and
demanding that the divergent parts cannot have the two-loop tensor structures
$b^{(2)}_1$ and $b^{(2)}_2$, we find
constraints that must be satisfied by the divergent parts:
\begin{eqnarray}
0 &=& 
 A^{\rm P}_Q(1,2,3,4) + A^{\rm P}_Q(3,4,2,1)
   - A^{\rm P}_Q(1,3,4,2) - A^{\rm P}_Q(4,2,3,1) \nonumber \\
&& \null 
 + A^{\rm NP}_Q(1,2,3,4) + A^{\rm NP}_Q(3,4,2,1) 
   - A^{\rm NP}_Q(1,3,4,2) - A^{\rm NP}_Q(4,2,3,1) 
\Bigr|_{D=5\, \rm div.} \,, \nn \\ 
0 & = & 
 A^{\rm P}_Q(1,4,2,3) + A^{\rm P}_Q(2,3,4,1) 
  - A^{\rm P}_Q(4,2,3,1) - A^{\rm P}_Q(1,3,4,2) \nonumber \\
&& \null
+ A^{\rm NP}_Q(1,4,2,3) + A^{\rm NP}_Q(2,3,4,1)
 - A^{\rm NP}_Q(4,2,3,1) - A^{\rm NP}_Q(1,3,4,2) 
\Bigr|_{D=5\, \rm div.} \,. 
\end{eqnarray}
Solving this for the divergent parts of two of the partial
amplitudes and plugging the solution into the supergravity expression
(\ref{TwoLoopSugraBCJ}), we immediately find that the corresponding
two-loop supergravity divergence in $D=5$ vanishes:
\begin{equation}
\mathcal{M}^{(2)}_{16+Q}(1,2,3,4) \Bigr|_{D=5\, \rm div.} = 0 \,.
\end{equation}
It is interesting that there is no need to impose the vanishing of the
contribution proportional to the one-loop color tensor $b_1^{(1)}$
to deduce this.  This demonstrates
that the cancellations that eliminate the $D=5$ divergence in the
two-loop four-point amplitude of half-maximal pure supergravity are
identical to the ones that eliminate forbidden color tensors from the
corresponding nonsupersymmetric pure Yang-Mills divergences. For
supergravity theories with more than $16$ supercharges, not only does
the divergence vanish for this reason, but it also vanishes because the $F^3$
operator (\ref{F3Operator}) in the corresponding super-Yang-Mills
theory is no longer a valid counterterm.

In $D = 6$, pure Yang-Mills has a two-loop divergence described by an
$F^4$ operator containing color factors not
appearing at tree level. (See \eqn{YMF4Operator}, but also containing
a two-loop color tensor.)  Feeding the $F^4$ counterterm of pure
Yang-Mills theory into the double-copy formula (\ref{TwoLoopSugraBCJ})
immediately shows that half-maximal supergravity diverges in
$D=6$.  Below we compute the explicit value of this divergence.

\subsubsection{Explicit cancellations in $D=5$}

We can see the supergravity divergence cancellation more directly in
a four-dimensional external subspace starting with the explicit values
of the $D=5$ pure Yang-Mills divergences computed in \App{YMD5App} for
identical external helicity states.  For Yang-Mills this external
helicity configuration is sufficient because it detects the divergence
generated by the $F^3$ operator. We note that the $({-}{+}{+}{+})$ external
helicity configuration is also divergent, but not the $({-}{-}{+}{+})$
case. This is because the allowed $F^3$ counterterm cannot generate
the latter helicity configuration.  The fact that the $D = 5$ pure
Yang-Mills amplitude with helicities $({-}{-}{+}{+})$ in the four-dimensional subspace
does not diverge at two loops immediately tells us that four-graviton
amplitudes in the four-dimensional subspace must also be finite: the
$({\pm}{+}{+}{+})$ graviton amplitude vanishes due to supersymmetric Ward
identities~\cite{SWI}, while the $({-}{-}{+}{+})$ graviton amplitude is finite
due to the lack of the corresponding Yang-Mills divergence.  On the
other hand, the presence of $({-}{+}{+}{+})$ or $({+}{+}{+}{+})$ pure Yang-Mills
divergences implies possible divergences in the supergravity theory with
one or two external scalars unless there are additional cancellations
beyond these helicity arguments, which, in fact, are present, as
described above.

To explicitly see these additional cancellations in the four-dimensional
external subspace, we use the results
for the planar and nonplanar contributions to the divergence 
given in the appendix, 
\begin{eqnarray}
A^\P(1^+,2^+,3^+,4^+)\Bigr|_{D=5\, \rm div.} &=& 
- i \, { \spb1.2\spb3.4 \over \spa1.2\spa3.4 }
s \,(D_s-2) \frac{\pi}{70 \epsilon} {1\over (4 \pi)^5}
              \,, \nn \\ 
A^\NP(1^+,2^+,3^+,4^+)\Bigr|_{D=5\, \rm div.} &=& 
i \, { \spb1.2\spb3.4 \over \spa1.2\spa3.4 }
 s \,(D_s-2) \frac{\pi}{70 \epsilon} {1\over (4 \pi)^5} \,,
 \end{eqnarray}
where we take the state-counting parameter to be $D_s = 5$ for the pure
Yang-Mills theory.

Plugging the above result back into the two-loop gravity amplitude
(\ref{TwoLoopSugraBCJ}), we immediately see that 
the divergences in the nonplanar contributions cancel with those
in the planar contributions, 
\begin{eqnarray}
\mathcal{M}^{(2)}(1^+,2^+,3^+,4^+) \Bigr|_{D=5 \, \rm div.}
&=&(D_s-2) \frac{\pi}{70 \epsilon} {1\over (4 \pi)^5}
\bigg(\frac{\kappa}{2}\bigg)^6stA^{\rm tree}_{Q=16}(1,2,3,4) \nn
\\
&& \hskip 1 cm \times
\bigg[s^2 \, { \spb1.2\spb3.4 \over \spa1.2\spa3.4 }\left(1-1+1-1\right)
+{\rm cyclic}(2,3,4)\bigg]\nn  \\
&=& 0\,,
\end{eqnarray}
valid for any external states in the graviton multiplet that are a
tensor product of the states of the $\NeqFour$ super-Yang-Mills
multiplet and identical-helicity gluons.  This explicit cancellation
highlights the fact that supergravity can be less divergent than the
component gauge-theory amplitudes because of cancellations between
planar and nonplanar contributions.

\subsection{Two loops and four points in $D=6$}

In $D=6$, on dimensional grounds one expects an $F^4$ counterterm in
pure Yang-Mills theory of the form in \eqn{YMF4Operator}, but with two-loop
color tensors.  As for the
one-loop $D=8$ case, the appearance of multiloop color tensors in the
gauge-theory divergence implies that the corresponding supergravity
divergences will not cancel.

In order to obtain the explicit value of the divergences, we follow
the same procedure as carried out in ref.~\cite{N4grav} for three-loop
$\NeqFour$ supergravity in $D=4$.  The ultraviolet divergences are
then extracted by expanding in external momenta and integrating, while
all subdivergences are subtracted integral by integral.

This construction yields the explicit form of the two-loop four-point
divergence for any external states in the graviton multiplet,
\begin{eqnarray}
\mathcal{M}^{(2)}(1,2,3,4) \Bigr|_{D=6\, \rm div.}\hskip -.4 cm\!  &=&
\!\frac{1}{(4\pi)^6}\left(\frac{\kappa}{2}\right)^6\!st A_{Q=16}^\tree(1,2,3,4)\bigg\{
\biggl(\frac{(D_s-6)(26-D_s)}{576\epsilon^2} +
            \frac{(19D_s-734)}{864 \epsilon}\biggr) \nn \\
&& \null\hskip 2 cm \times 
\bigg[ s\, (F_1 F_2)(F_3 F_4) + t\, (F_1 F_4)(F_2 F_3) + u\, (F_1 F_3)(F_2 F_4) 
\bigg]\nn \\
&&\null 
+\frac{(48D_s-1248)}{864\epsilon}\bigg[u \, (F_1 F_2 F_3 F_4)
  + t \, (F_1 F_3 F_4 F_2)+s \, (F_1 F_4 F_2 F_3)\bigg]\bigg\}\,,\nn \\
\end{eqnarray}
including the subtraction of one-loop subdivergences that appear for
$D_s \not = 6$.  These subdivergences come from extra states that
circulate in the loop when $D_s \not = 6$.  For pure half-maximal
supergravity (where the state-counting parameter is $D_s=6$), the
$1/\eps^2$ divergence vanishes as expected since, as discussed in
\sect{OneloopUVSection}, there are no one-loop subdivergences in pure
half-maximal supergravity.

We can simplify the expression for the divergences in a
four-dimensional external subspace using spinor helicity.  For example,
for four external gravitons with helicity
configuration (${-}{-}{+}{+}$) we have
\begin{equation}
\mathcal{M}^{(2)}(1^-,2^-,3^+,4^+)
= -\frac{i}{(4\pi)^6}\, \left(\frac{\kappa}{2}\right)^6 
 \Bigl(\frac{(D_s-6)(26-D_s)}{576 \eps^2} 
                                 + \frac{19 D_s-734}{864\eps} \Bigr)
  \, s\langle12\rangle^4[34]^4 \,,
\label{TwoLoopD6Helicity}
\end{equation}
for the one-loop-subtracted result.
Among the $(F_i F_j)(F_k F_l)$ terms on the pure Yang-Mills side, only $(F_1 F_2)(F_3 F_4)$ gives a nonvanishing contribution, while the contributions of the $(F_i F_j F_k F_l)$ terms cancel among themselves.  We note that the expression
(\ref{TwoLoopD6Helicity}) has the helicity structure and dimensions
of a $D^2R^4$ counterterm.

\subsection{Two loops and five points in $D=5$}

We now turn our attention to five points.  While the previous discussion
rules out an $R^4$ divergence in $D=5$, one may worry about a counterterm of the
form $\phi R^4$ and its supersymmetric completion, which
would lead to a divergence at five points.  However, from the SO(1,1)
duality symmetry obeyed by half-maximal supergravity in
$D=5$~\cite{D5N4Sugra}, we know that $\phi R^4$ is not a valid
counterterm because it is not invariant under the $\phi \rightarrow \phi
+ v$ shift symmetry. Nevertheless, it is interesting to see how the
potential divergence cancels from the double-copy vantage point.

For the two-loop five-point amplitudes, the numerators of maximal
super-Yang-Mills theory depend on loop momenta~\cite{BCJ5Point}.  This
complicates the analysis of the corresponding half-maximal
supergravity theory, though it is straightforward to work out the
divergences in $D=5$ following the procedure of ref.~\cite{N4grav}.

\begin{figure}[tbh]
\centering
\subfigure[]{\epsfxsize 1.7 truein \epsfbox{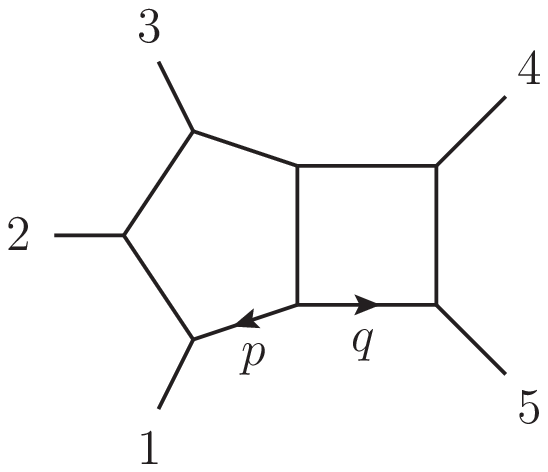}}
\hspace{1cm}
\subfigure[]{\epsfxsize 1.7 truein \epsfbox{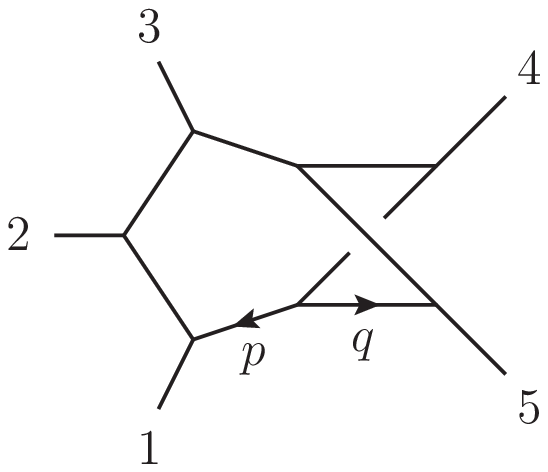}}
\hspace{1cm}
\subfigure[]{\epsfxsize 1.7 truein \epsfbox{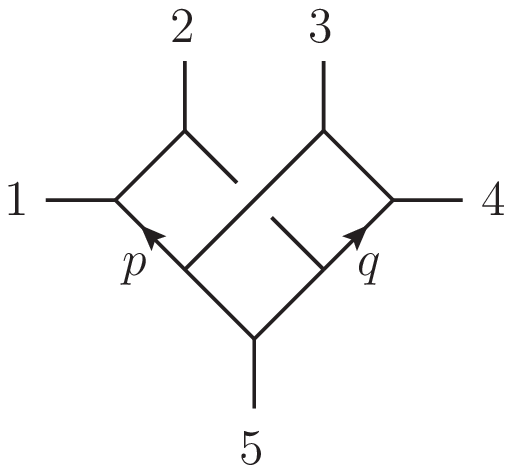}}
\\
\subfigure[]{\epsfxsize 1.7 truein \epsfbox{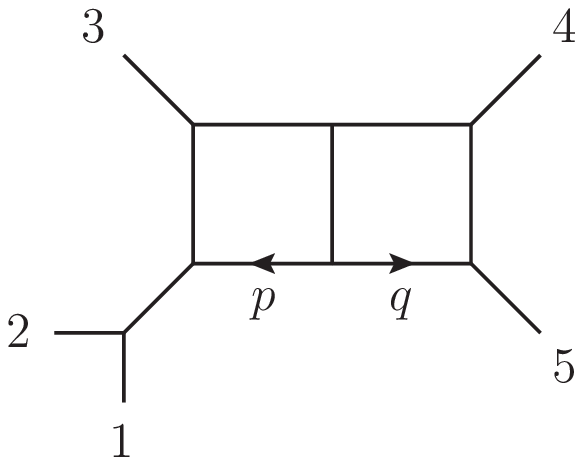}}
\hspace{1cm}
\subfigure[]{\epsfxsize 1.7 truein \epsfbox{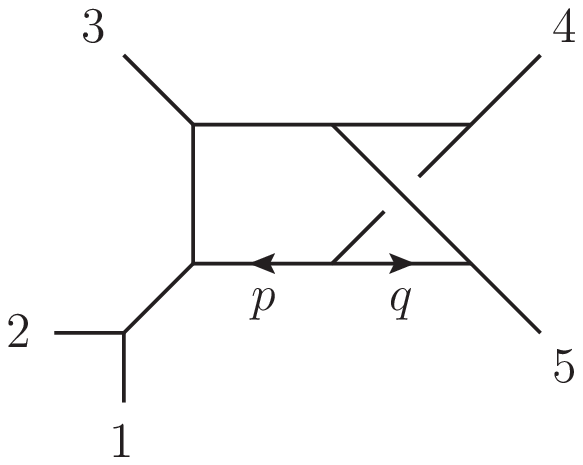}}
\hspace{1cm}
\subfigure[]{\epsfxsize 1.7 truein \epsfbox{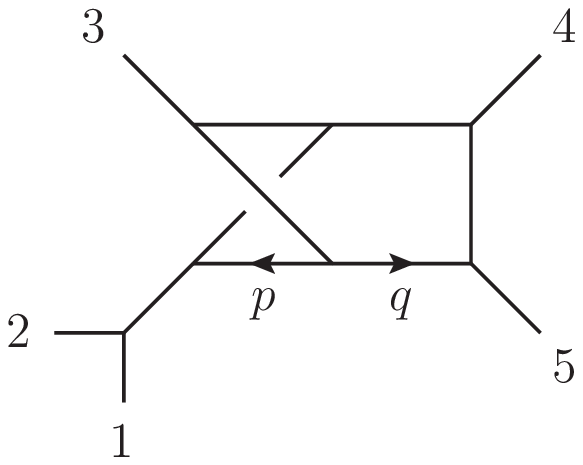}}
\caption[a]{Diagrams contributing to the five-point two-loop amplitude of 
maximal super-Yang-Mills theory. From ref.~\cite{BCJ5Point}.
}
\label{TwoLoopFigure}
\end{figure}

\begin{table*}[t]\caption{The numerator factors of the graphs in
\fig{TwoLoopFigure}. The first column indicates the integral, the
second column the numerator factor for maximal $\NeqFour$
super-Yang-Mills five-gluon MHV amplitudes, where 
the external momenta and states live in a four-dimensional subspace.
From ref.~\cite{BCJ5Point}.
\label{NumeratorTable} }
\vskip .4 cm
\begin{center}
\begin{tabular}{||c|c||}
\hline
 ${\cal I}^{(x)}$ & maximal super-Yang-Mills numerator \\
\hline
\hline
(a),(b) & 
$\frac{1}{4} \Bigl(\gamma_{12} (2 s_{45}- s_{12} + \tau_{2p} - \tau_{1p})+ \gamma_{23} (s_{45}  + 2 s_{12} - \tau_{2p} + \tau_{3p})   $\\
&$\null 
\hskip 2.3cm + 2 \gamma_{45} (\tau_{5p} - \tau_{4p})+ \gamma_{13} (s_{12}  + s_{45} - \tau_{1p} + \tau_{3p}) \Bigr)$
\\
\hline 
(c) &
$\frac{1}{4} \Bigl( \gamma_{15} (\tau_{5p} - \tau_{1p})+ \gamma_{25} (s_{12} - \tau_{2p} + \tau_{5p}) + \gamma_{12} (s_{34} + \tau_{2p} - \tau_{1p} + 2 s_{15} + 2 \tau_{1q} -  2 \tau_{2q}) $\\
&$\null 
~~+ \gamma_{45}  (\tau_{4q} - \tau_{5q}) - \gamma_{35} (s_{34} - \tau_{3q} + \tau_{5q}) + \gamma_{34} (s_{12} + \tau_{3q} - \tau_{4q} + 2 s_{45} + 2 \tau_{4p} - 2 \tau_{3p} ) \Bigr)$ 
\\
\hline
(d)-(f) & $  \gamma_{12} s_{45}-\frac{1}{4} \Bigl(2 \gamma_{12} + \gamma_{13} - \gamma_{23}\Bigr) s_{12} $
 \\
\hline
\end{tabular}
\end{center}
\end{table*}

Once again we employ the double-copy construction (\ref{DoubleCopy})
to obtain the results for pure half-maximal supergravity.  In
ref.~\cite{BCJ5Point}, a form of the maximal super-Yang-Mills
amplitude that satisfies BCJ duality was found for any internal
dimension with the external states restricted to a four-dimensional
subspace.  We employ this here for pure gluon amplitudes. In the
double copy this gives us access to all states obtained by tensoring
two gluon states in the subspace. The graphs with nonvanishing
numerators for maximal super-Yang-Mills are shown in
\fig{TwoLoopFigure}, and the corresponding numerators are in
\tab{NumeratorTable}~\cite{BCJ5Point}.  Since there is no need to have
a BCJ form in the second copy, we follow ref.~\cite{N4grav} and use
ordinary Feynman-gauge Feynman diagrams on the pure Yang-Mills side to
generate a set of suitable numerators. (See ref.~\cite{N4grav} for a
description of this procedure.)  While using Feynman diagrams as a
starting point is not efficient, enormous simplifications arise from
the fact that we do not need contributions corresponding to those with
vanishing numerators on the maximal super-Yang-Mills side.  Unlike the
cases covered earlier, the maximal ($Q=16$) super-Yang-Mills two-loop
five-point numerators contain loop momenta and therefore cannot be
pulled out of the integral.

A generic integral for a graph in \fig{TwoLoopFigure} is of the form,
\begin{equation}
I^{(x)}=\int\frac{d^Dp}{(2\pi)^D}\frac{d^Dq}{(2\pi)^D}\frac{n_{Q=16}^{(x)}(1,2,3,4,5;p,q)n_{Q=0}^{(x)}(1,2,3,4,5;p,q)}{\prod_{\alpha_{(x)}}{l^2_{\alpha_{(x)}}}},
\end{equation}
where $n_{Q=16}$ denotes the maximal super-Yang-Mills numerators specified 
in \tab{NumeratorTable} and $n_{Q=0}$ the pure Yang-Mills numerator found 
via Feynman rules.  Including the symmetry factors, the gravity amplitude 
is then given by
\begin{equation}
\mathcal{M}_{Q+16}^{(2)} (1,2,3,4,5) = -i\left(\frac{\kappa}{2}\right)^7
\sum_{\mathcal{S}_5}
\left(\frac{1}{2}I^{\rm (a)}+\frac{1}{4}I^{\rm (b)}+\frac{1}{4}I^{\rm (c)}
+\frac{1}{2}I^{\rm (d)}+\frac{1}{4}I^{\rm (e)}+\frac{1}{4}I^{\rm (f)}\right)\,,
\end{equation}
where the sum $\mathcal{S}_5$ is over all permutations of external legs.

We carry out the extraction of the potential ultraviolet divergences
exactly as in ref.~\cite{N4grav}, to which we refer the reader. In
brief, we extract the ultraviolet divergences by expanding the
external momenta~\cite{MarcusSagnotti}, as has been recently carried
out in various determinations of ultraviolet divergences in
super-Yang-Mills theory and
supergravity~\cite{GravityThree,CompactThree,GravityFour,Neq44np,ck4l,N4grav,Neq54np}.
The resulting vacuum integrals are reduced to a basis using
FIRE~\cite{Fire}, giving integrals that are straightforward to
evaluate.  In $D=5$, there are no subdivergences to subtract,
simplifying the construction compared to ref.~\cite{N4grav}.

\begin{table}[t]
  \begin{center}
\begin{tabular}[t]{|c|c|}
\hline
Graph & $(\text{divergence})/(i \gamma_{12}\,\pol_1\cdot\pol_3\,\pol_4\cdot\pol_5\,k_1\cdot\pol_2\,s_{12}/(4\pi)^5)$ \\
\hline
(a) & $\frac{-64497+925 D_s}{362880\sqrt{2}}\frac{1}{\epsilon}$ \\
(b) & $\frac{820641-149788 D_s}{1451520\sqrt{2}}\frac{1}{\epsilon}$ \\
(c) & $\frac{-27555+8116 D_s}{80640\sqrt{2}}\frac{1}{\epsilon}$ \\
(d) & $\left(\frac{20605+912 D_s}{53760\sqrt{2}}+\frac{-38+D_s}{240\sqrt{2}}\frac{s_{14}}{s_{13}}+\frac{655-161 D_s}{1680\sqrt{2}}\frac{s_{23}}{s_{13}}+\frac{-5171-148 D_s}{6720\sqrt{2}}\frac{s_{24}}{s_{13}}\right)\frac{1}{\epsilon}$ \\
(e) & $\left(\frac{-71986+4511 D_s}{241920\sqrt{2}}+\frac{935+6 D_s}{6720\sqrt{2}}\frac{s_{14}}{s_{13}}+\frac{-907+342 D_s}{6720\sqrt{2}}\frac{s_{23}}{s_{13}}+\frac{27859+844 D_s}{60480\sqrt{2}}\frac{s_{24}}{s_{13}}\right)\frac{1}{\epsilon}$ \\
(f) & $\left(\frac{-31847-8615 D_s}{241920\sqrt{2}}+\frac{129-34 D_s}{6720\sqrt{2}}\frac{s_{14}}{s_{13}}+\frac{-1713+302 D_s}{6720\sqrt{2}}\frac{s_{23}}{s_{13}}+\frac{2335+61 D_s}{7560\sqrt{2}}\frac{s_{24}}{s_{13}}\right)\frac{1}{\epsilon}$ \\
\hline
  \end{tabular}
  \end{center}
\caption{The graph-by-graph divergent coefficients of the term
  containing the factor $i \gamma_{12}\,\epsilon_1\cdot\epsilon_3\,
  \epsilon_4\cdot\epsilon_5\,k_1\cdot\epsilon_2\,s_{12}/(4\pi)^5$ for
  the two-loop five-point half-maximal supergravity amplitude in
  $D=5$. As discussed in the text we have reduced each expression to a
  set of terms independent under momentum conservation and spinor
  identities.  Each expression in the table includes a permutation sum
  over external legs, with the symmetry factor appropriate to the
  indicated graph.  The sum of contributions over all graphs vanishes
  for any value of the state-counting parameter $D_s$; all other
  divergent terms amplitude similarly cancel. }
\label{TwoLoop5PtTable}
\end{table}

As was the case at four points, we find the divergence to vanish:
\begin{equation}
\mathcal{M}^{(2)}_{Q=16}(1,2,3,4,5) \Bigr|_{D=5\, \rm div.} = 0 \,.
\end{equation}
This result is valid for any states obtained by tensoring a pair of
gluon states restricted to a four-dimensional subspace. 
The cancellation of the divergence between graphs
for one independent term is shown in Table \ref{TwoLoop5PtTable}.
Each row gives the divergent coefficient of the term
$i\gamma_{12}\,\pol_1\cdot\pol_3\,\pol_4\cdot\pol_5\,
k_1\cdot\pol_2\,s_{12}/(4\pi)^5$ from the indicated graph in
\fig{TwoLoopFigure}. This includes the sum over permutations of
external legs.  We have applied momentum conservation as well as taken
a basis of six $\gamma_{ij}$. Our choice is to eliminate $k_5$ via
\begin{equation}
k_5\cdot\pol_i = -(k_1+k_2+k_3+k_4)\cdot \pol_i \,, \hskip 1 cm 
k_4\cdot\epsilon_5 = -(k_1+k_2+k_3)\cdot\epsilon_5 \,, \hskip 1 cm 
k_i \cdot \pol_i = 0\,.
\end{equation}
We use the five independent Mandelstam invariants $s_{12}, s_{13},
s_{14}, s_{23}$ and $s_{24}$.  The six independent numerator factors
are $\gamma_{12}, \gamma_{13}, \gamma_{14}, \gamma_{23}, \gamma_{24}$
and $\gamma_{34}$.  This gives a total of thirty monomials
$\gamma_{ij}s_{kl}$; however, as explained in ref.~\cite{BCJ5Point},
there are actually only twenty-five independent ones due to nontrivial
additional relations amongst them.  We have used this fact to
eliminate the following monomials from our graph-by-graph results:
\begin{equation}
\gamma_{12}s_{14}\,, \hskip 1 cm 
\gamma_{12}s_{23}\,, \hskip 1 cm 
\gamma_{13}s_{12}\,, \hskip 1 cm 
\gamma_{13}s_{13}\,, \hskip 1 cm 
\gamma_{34}s_{24}\,.
\end{equation}
After reducing to this basis (or any similar one), all divergences
completely cancel in a manner similar to the cancellation obtained by
summing the contributions in \tab{TwoLoop5PtTable}. It is interesting
that this cancellation is independent of the state-counting parameter
$D_s$.


\section{Conclusions and Outlook}
\label{ConclusionSection}

In a previous paper~\cite{N4grav}, we proved that at three loops in
$\NeqFour$ supergravity an $R^4$ counterterm---valid under all
currently known supersymmetry and duality
constraints~\cite{VanishingVolume}---has vanishing coefficient.  In
the present paper, we analyzed the simpler two-loop case of pure
half-maximal supergravity in $D=5$, which is believed to have a valid
counterterm under all known supersymmetry and duality constraints.
However, using the double-copy structure, we showed that the
corresponding divergences completely cancel.  Indeed we found that
there are no four-point divergences in $D<8$ at one loop and in $D<6$
at two loops, and we linked these cancellations to similar ones
occurring in corresponding pure Yang-Mills amplitudes that prevent
forbidden color structures from appearing in divergences.  We also
reached the same conclusions for the five-point amplitudes that we
analyzed at one and two loops.  This link between gravity and gauge
theory is consistent with previous explicit calculations showing that
ultraviolet divergences of supergravity theories can bear a strong
resemblance to those of corresponding gauge theories, not only in
their general structure but in their details~\cite{Neq44np,ck4l}.

For the half-maximal supergravity one- and two-loop four- and
five-point cases studied here, when divergences of the corresponding
pure-Yang-Mills amplitudes contain color structures other than the
tree ones, then the supergravity amplitudes also diverge.  In $D=8$ and
at two loops in $D=6$ the pure Yang-Mills divergences have such color
factors so the half-maximal supergravity amplitudes also diverge. In
lower dimensions, only tree color tensors appear, so the corresponding
supergravity amplitudes are finite.  Using the double-copy formula we
also presented explicit expressions for the valid supergravity
divergences in terms of Yang-Mills ones.  

The above results are suggestive of a strong link between the
divergences of the two theories when the number of loops or legs
increases.  With larger numbers of loops or legs, loop momenta can
appear in both gauge-theory numerator factors of certain diagrams in
the double-copy formula.  This makes it is more difficult to directly
tie the integrated divergence properties of supergravity theories to
gauge theories.  Nevertheless, it is rather striking that the
finiteness of the three-loop four-point $\NeqFour$ supergravity
amplitude~\cite{N4grav} is correlated with the lack of multiloop color
tensors in the corresponding pure Yang-Mills divergences, suggesting a
general pattern.  Similarly, we found nontrivial cancellations in
$D=5$ five-point two-loop amplitudes of half-maximal supergravity,
even though both gauge-theory copies have loop momenta in their
numerators.  An obvious conjecture is that the pattern continues to
higher loops, with divergences possible in $(Q+16)$-supercharge
supergravity only when the divergences of corresponding
$Q$-supercharge gauge theory contain independent color tensors other
than tree ones.  In $D=4$ this would imply ultraviolet finiteness of
pure ${\cal N} \ge 4$ supergravity.

In order to test this conjecture and to guide future studies, it is,
of course, crucially important to carry out further explicit studies
of divergences with larger numbers of loops or legs.  In particular, a
computation of the five-loop four-point divergence in $\NeqEight$
supergravity should be within reach~\cite{Neq54np}, now that the
corresponding $\NeqFour$ super-Yang-Mills integrand has been
obtained~\cite{Neq54np} (although not in a BCJ format). The
calculation of the four-loop four-point divergence of $\NeqFour$
supergravity in $D=4$ is also doable with the procedure of
ref.~\cite{N4grav} since the BCJ form of the corresponding $\NeqFour$
super-Yang-Mills amplitude required by the double-copy formula is
known~\cite{ck4l}.

There are a number of other obvious directions for future research.  A
key issue is to find the extent to which supersymmetry and duality
symmetries by themselves can be used to place restrictions on counterterms
corresponding to the results described here.  Very interestingly, the
potential two-loop four-point $D=5$ counterterm does appear to be a
duality satisfying full superspace integral of a density (which itself
is not duality invariant) so such an explanation would be
nontrivial~\cite{BossardHoweStelle5D}.  It would be interesting to see
if any of the recent developments in tree-level gravity
amplitudes~\cite{RecentTreeGravity} can shed any light on the
nontrivial ultraviolet cancellations we see at loop level.

In summary, in this paper we linked the divergences of half-maximal
supergravity to those of pure Yang-Mills theory.  In particular, for
the $D=5$ two-loop four-point amplitudes of half-maximal supergravity,
the divergences vanish via the same cancellations that remove
forbidden color factors from the divergences of corresponding pure
Yang-Mills amplitudes.  This case was particularly simple to analyze
because the maximal super-Yang-Mills numerators used in the
double-copy construction are independent of loop momenta.  The next
challenge is to fully unravel the ultraviolet cancellations implied by
the double-copy structure at higher-loop orders.

\vskip .2 cm 
\subsection*{Acknowledgments}
We thank G.~Bossard and K.~Stelle for many important discussions
motivating the present paper and for informing us of the content of
their forthcoming paper on half-maximal
supergravity~\cite{BossardHoweStelle5D}.  We thank J.~J.~Carrasco,
L.~Dixon,  H.~Johansson and R.~Roiban for many helpful discussions. as well
as for important comments on the manuscript.  We also thank S.~Ferrara
and P.~Vanhove for helpful discussions.  This research was
supported by the US Department of Energy under contract
DE--FG03--91ER40662.

\appendix
\section{Two-loop pure Yang-Mills divergence in $D=5$}
\label{YMD5App}

In this appendix, we explicitly compute the $D=5$ divergence of the
two-loop pure Yang-Mills four-point amplitude.  The counterterm in
this case is the $F^3$ operator (\ref{F3Operator}).

To simplify the analysis we restrict ourselves to a four-dimensional
external subspace.  In this subspace, the operator generates
nonvanishing contributions to the $({+}{+}{+}{+})$ helicity states.  The
all-plus helicity two-loop integrand in Yang-Mills was given in
ref.~\cite{TwoLoopAllPlus} in a form valid for arbitrary internal
dimensions.  Here we integrate this expression in $D=5 -2 \eps$
to obtain the explicit form of the ultraviolet divergence.
We then use this expression to explicitly confirm our more general
discussion of the cancellations of the divergences in $D=5$
half-maximal supergravity.

\begin{figure}
\begin{center}
\includegraphics[scale=0.5]{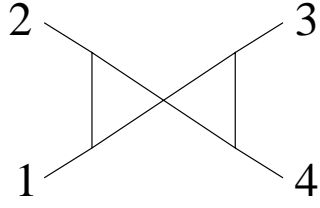}
\caption{The bow-tie integral. }
\label{ParentsFigure2}
\end{center}
\end{figure}

The unintegrated form of the pure Yang-Mills amplitude with 
identical external helicities is~\cite{TwoLoopAllPlus}
\begin{eqnarray}
A^\P(1^+,2^+,3^+,4^+)
& =& \! i \, { \spb1.2\spb3.4 \over \spa1.2\spa3.4 }
 \biggl\{
 s \, \I_4^\P (s,t) + 4 (D_s-2) \, \I_4^\bowtie[(\mud_p^2 + \mud_q^2) 
                \, (\mud_p \cdot \mud_q) ] (s) \nn \\
&& \hskip 2.2 cm
+ {(D_s-2)^2 \over s} \, \I_4^\bowtie\Bigl[ 
      \mud_p^2 \, \mud_q^2 \, ( (p+q)^2 + s ) \Bigr] (s,t)
              \biggr\} \,, \hskip 1.5 cm  \label{ppppPlanar} \nn \\
A^\NP(1^+,2^+,3^+,4^+)
&= &  i \, { \spb1.2\spb3.4 \over \spa1.2\spa3.4 } \, s \,
\I_4^\NP (s,t)\,,
\label{ppppNonPlanar}
\end{eqnarray}
where $D_s$ is the state-counting parameter~\cite{FDH}.  In pure
half-maximal supergravity we take $D_s = 5$. Here, the external
kinematics are four-dimensional, while the loop momenta are in
$D=5-2\epsilon$, and $(\lambda_p,\lambda_q)$ are the
$(D-4)$-dimensional components of the two-loop momenta. The planar and
nonplanar double-box integrals are defined as
\begin{eqnarray}
&&\hskip -.3 cm  \I_4^\P (s,t) 
 \equiv \! \int {d^{D}p\over (2\pi)^{D}} \, {d^{D}q\over (2\pi)^{D}}\,
 { (D_s-2) ( \mud_p^2 \, \mud_q^2 
         + \mud_p^2 \, \mud_{p+q}^2  + \mud_q^2 \, \mud_{p+q}^2 ) 
       + 16 \Bigl[ (\mud_p \cdot \mud_q)^2 - \mud_p^2 \, \mud_q^2 \Bigr] 
                     \over 
     p^2\, q^2\, (p+q)^2 (p - k_1)^2 \,(p - k_1 - k_2)^2 \,
        (q - k_4)^2 \, (q - k_3 - k_4)^2 } \,,  \nn\\
\nn \\
&&\hskip -.3 cm \I_4^\NP (s,t) \nn  \\
&& \hskip 1.2 cm 
\equiv \int \! {d^{D} p \over (2\pi)^{D}} \,
        {d^{D} q \over (2\pi)^{D}}\,
{(D_s-2) ( \mud_p^2 \, \mud_q^2 
         + \mud_p^2 \, \mud_{p+q}^2  + \mud_q^2 \, \mud_{p+q}^2 ) 
       + 16 \Bigl[ (\mud_p \cdot \mud_q)^2 - \mud_p^2 \, \mud_q^2 \Bigr] \over p^2\, q^2\, (p+q)^2 \,
         (p-k_1)^2 \,(q-k_2)^2\,
   (p+q+k_3)^2 \, (p+q+k_3+k_4)^2}\,, \nn  \\
\label{PlanarNonPlanarInt}
\end{eqnarray}
with corresponding diagrams shown in \fig{DoubleBoxFigure}. The `bow-tie'
double-triangle integrals, displayed in \fig{ParentsFigure2}, are defined
as
\begin{eqnarray}
&& \I_4^{\bowtie} [{\cal P}  (\mud_i, p,q,k_i)] (s) \nn \\
&& \hskip 2 cm \null
  \equiv \int {d^{D}p\over (2\pi)^{D}} \,
 {d^{D}q\over (2\pi)^{D}}\,
 { {\cal P} (\mud_i, p,q,k_i) \over 
     p^2\, q^2\, (p - k_1)^2 \,(p - k_1 - k_2)^2 \,
        (q - k_4)^2 \, (q - k_3 - k_4)^2 }\,. \hskip 1 cm 
\label{BowTieInt} 
\end{eqnarray}

We now compute the divergent parts of the integrals.  
In five dimensions, there are no infrared divergences so all divergences 
are ultraviolet in nature.

%

The bow-tie integrals are finite in five dimensions:
\begin{eqnarray}
\I_4^\bowtie[\mud_p^2 \mud_q^2](s) &=& {\pi^3s\over 576} \, 
   {1\over (4 \pi)^5}\,, \cr
\I_4^\bowtie[\mud_p^2 \mud_q^2 (p+q)^2](s,t) 
&=& {\pi^3s(2t-15s)\over 18432} {1\over (4\pi)^5}  \,,\nn \\
\I_4^\bowtie[\mud_p^2 \, (\mud_p \cdot \mud_q)](s) &=& 0\,,
\label{BowTieInts}
\end{eqnarray}
thus the ultraviolet divergence comes solely from the double-box integrals.

Using Schwinger parameters, we write the planar double-box 
integral in eq.~(\ref{PlanarNonPlanarInt}) with constant numerator as 
\begin{equation}
 \I_4^\P [1] (s,t)=\prod^7_{i=1}\int_0^\infty dt_{i} \left[\Delta_\P(T)\right]^{-\frac{D}{2}} \exp\left[ -\frac{ Q_\P(s,t,t_i)}{\Delta_\P(T)}\right] \,,
\label{feynparam}
\end{equation}
where
\begin{equation}
 \Delta_\P(T) = (T_pT_q+T_pT_{pq}+T_qT_{pq})\,,
\label{Deltadenom}
\end{equation}
with 
\begin{equation}
T_p = t_3+t_4+t_5,\;\;\;T_q=t_1+t_2+t_7,\;\;\;T_{pq}=t_6 \,. 
\end{equation}
As the subscripts indicate, $T_p,T_q$ and $T_{pq}$ are the sum of Schwinger parameters whose corresponding propagators contain loop momenta $p,q$ and $p+q$ respectively. Finally, we also have
\begin{equation}
  Q_\P(s,t,t_i) = -s \, \bigg( t_1 t_2 T_p + t_3 t_4 T_q 
                       + t_6 (t_1+t_3)(t_2+t_4) \bigg) 
              -t \, t_5 t_6 t_7 \,. 
\label{Qnumer}
\end{equation}
The effects of $\lambda^2_p,\lambda^2_q$ and $\lambda^2_{p+q}$ in 
the numerators are derived by taking derivatives on
\begin{equation}
\int d\lambda_p^{1-2\epsilon}d\lambda_q^{1-2\epsilon} \exp\left[ -T_p\lambda^2_p-T_q\lambda^2_q-T_{pq}\lambda^2_{p+q}\right]\propto \left[\Delta_\P(T)\right]^{-\frac{1}{2}+\epsilon}\,,
\end{equation}
with respect to $T_p,T_q$ and $T_{pq}$. This leads to the following extra factors, for example, to be inserted in the integrand of eq.~(\ref{feynparam})\,,
\begin{eqnarray}
 \lambda^4_p    &\rightarrow& (\epsilon-\frac{1}{2})(\epsilon-\frac{3}{2})\frac{(T_{pq}+T_q)^2}{\Delta^2_\P(T) } \,, \nn \\
 \lambda^2_p\lambda^2_{p+q}   &\rightarrow&  \frac{(\epsilon-\frac{1}{2})^2}{\Delta_\P(T) }+\frac{(\epsilon-\frac{1}{2})(\epsilon-\frac{3}{2})T^2_{q}}{\Delta^2_\P(T) } \,. 
\label{extra}
\end{eqnarray}      
We account for the extra factors of $\Delta^a_\P(T)$ by shifting 
the dimension $D\rightarrow D-2a$.  We now change six of the Schwinger parameters to Feynman parameters such that the delta-function constraint on the Feynman parameters is $\sum_{i\neq6}\alpha_i=1$.  We then have
\begin{equation}
\I_4^\P [{\cal P}  (\mud_p, \mud_q)] (s,t)=\Gamma[7-D+\gamma]
\int_0^{\infty}\hspace{-.1cm}d\alpha_6\prod_{i\neq6}\int_0^1 \hspace{-.1cm}d\alpha_{i} \delta\left(\hspace{-.05cm}1-\sum_{i\neq6}\alpha_i\right)\frac{\left[\Delta_\P(T)\right]^{7-\frac{3D}{2}+\gamma}}{ \left[ Q_\P(s,t,\alpha_i)\right]^{7-D+\gamma}}D(\alpha_i) \,,
\label{FeynPa}
\end{equation}
where $D(\alpha_i)$ are the extra factors in \eqn{extra}, with
$t_i\rightarrow \alpha_i$. If the extra factors in \eqn{extra}
depend on $T_p,T_q$ and $T_{pq}$, then $\gamma=2$; otherwise, we have
$\gamma=0$. Following Smirnov~\cite{Smirnov}, we perform a change of variables that imposes the delta-function constraint:
\begin{eqnarray}
&&\alpha_1=\beta_1\xi_3\,, \hskip .6 cm 
\alpha_2=(1-\xi_5)(1-\xi_4)\,,     \hskip .6 cm 
\alpha_3 =\beta_2\xi_1\,, \hskip .6 cm
\alpha_4=\xi_5(1-\xi_2)\,, \nn \\
&&\alpha_5=\beta_2(1-\xi_1)\,, \hskip .6cm
\alpha_7=\beta_1(1-\xi_3)\,, \hskip .6cm
\beta_1=(1-\xi_5)\xi_4\,, \hskip .6cm
\beta_2=\xi_5\xi_2\,.
\end{eqnarray}
The parameters can then be straightforwardly integrated
to obtain a Mellin-Barnes representation, and explicit integration gives 
\begin{eqnarray}
  \I_4^\P [\lambda_p^2\lambda_q^2] &=& \frac{\pi}{70 \epsilon} {1\over (4 \pi)^5}+\mathcal{O}(\epsilon^0) \,, \nn \\
  \I_4^\P [\lambda_p^2\lambda_{p+q}^2] &=& -\frac{\pi}{70 \epsilon} {1\over (4 \pi)^5}+\mathcal{O}(\epsilon^0) \,, 
\cr
\I_4^\P [\lambda_p^4] &=& -\frac{\pi}{70 \epsilon} {1\over (4 \pi)^5}+\mathcal{O}(\epsilon^0) \,, \cr
  \I_4^\P [\lambda_{p+q}^4]  &=& \mathcal{O}(\epsilon^0) \,.
\end{eqnarray}
Inserting these results into eq.~(\ref{ppppPlanar}), the all-plus helicity planar amplitude is
\begin{equation}
A^\P(1^+,2^+,3^+,4^+)  = i \, { \spb1.2\spb3.4 \over \spa1.2\spa3.4 }
 \biggl\{ -s \,(D_s-2) \frac{\pi}{70 \epsilon} {1\over (4 \pi)^5}
  +\mathcal{O}(\epsilon^0) \biggr\} \,.
\label{ppppPlanarR}
\end{equation}

The evaluation of the nonplanar double-box integrals follows the same steps as the planar ones, with $\Delta_\NP(T)$ taking the same form as 
$ \Delta_\P(T)$, but now identifying: 
\begin{equation}
T_p = t_1+t_2,\;\;\;T_q=t_3+t_4,\;\;\;T_{pq}=t_5+t_6+t_7 \,.
\end{equation}
Similarly, we also have
\begin{equation}
  Q_\NP(s,t,u,t_i)  = -s \, \bigl( t_1 t_3 t_5 + t_2 t_4 t_7 
              + t_5 t_7 (T_p+T_q) \bigr)
              - t \, t_2 t_3 t_6 -u \, t_1 t_4 t_6 \,. 
\end{equation}
However, here we find it advantageous to change only four Schwinger parameters to Feynman parameters.  Performing this change gives
\begin{eqnarray}
 \I_4^\NP [{\cal P}  (\mud_p, \mud_q)] &=&\Gamma[7-D+\gamma] \\
&& \null \times 
\prod^7_{i=5}\int_0^\infty d\alpha_{i} \prod^4_{j=1}
\int_0^1 d\alpha_{j} \delta\left(1-\sum_{i=1}^4\alpha_i\right)
\frac{\left[\Delta_\NP(T)\right]^{7-\frac{3D}{2}+\gamma}}
     {\left[ Q_\NP(s,t,u,\alpha_i)\right]^{7-D+\gamma}}
 D(\alpha_i) \,.\nn \hskip .5 cm
\end{eqnarray}
The delta-function constraint can be imposed via further redefinition:
\begin{equation}
\alpha_1=\xi_3(1-\xi_1)\,, \hskip .8 cm 
\alpha_2=\xi_3\xi_1\,,     \hskip .8 cm 
\alpha_3 =(1-\xi_3)(1-\xi_2)\,, \hskip .8 cm
\alpha_4=(1-\xi_3)\xi_2\,.
\end{equation}
The parameters can once again be straightforwardly integrated, and we arrive at
\begin{eqnarray}
\I_4^\NP [\lambda_p^2\lambda_q^2] &=& -\frac{\pi}{42 \epsilon} {1\over (4 \pi)^5}+\mathcal{O}(\epsilon^0) \,,\nn \\
  \I_4^\NP [\lambda_p^2\lambda_{p+q}^2] &=& \frac{2\pi}{105 \epsilon} {1\over (4 \pi)^5}+\mathcal{O}(\epsilon^0) \,, \nn \\
\I_4^\NP [\lambda_p^4] &=& \mathcal{O}(\epsilon^0) \,, \nn \\
  \I_4^\NP [\lambda_{p+q}^4] &=&
  \frac{\pi}{35 \epsilon} {1\over (4 \pi)^5}+\mathcal{O}(\epsilon^0) \,.
\end{eqnarray}
Inserting these results into eq.~(\ref{ppppNonPlanar}), the all-plus
helicity nonplanar amplitude is given by
\begin{equation}
A^\NP(1^+,2^+,3^+,4^+) = i \, { \spb1.2\spb3.4 \over \spa1.2\spa3.4 }
 \biggl\{ s \,(D_s-2) \frac{\pi}{70 \epsilon} 
          {1\over (4 \pi)^5}+\mathcal{O}(\epsilon^0)  \biggr\} \,. 
\label{ANP}
\end{equation}
We use the results for the two-loop divergences in
\eqns{ppppPlanarR}{ANP} in \sect{TwoloopUVSection} to explicitly demonstrate
the cancellation of the corresponding divergence of $D=5$ half-maximal
supergravity.


\end{document}